\documentclass[
10pt,A4paper,twocolumn,aps,pra, superscriptaddress,longbibliography,nofootinbib,amsmath,amssymb,
]{revtex4-2}

\usepackage[utf8]{inputenc} 
\usepackage[T1]{fontenc}    
\usepackage{lmodern}        
\usepackage{microtype} 
\usepackage{graphicx} 
\usepackage{enumerate}
\usepackage{amsmath}
\usepackage{amsfonts}
\usepackage{amssymb}
\usepackage{bm} 
\usepackage{braket}
\usepackage{mathtools}
\usepackage{enumitem}
\usepackage[colorlinks=true, linkcolor=blue, citecolor=blue, urlcolor=blue, pdfauthor={Besedin I., Kerschbaum M., et. al.}, pdftitle={Realizing Lattice Surgery on Two Distance-Three Repetition Codes with Superconducting Qubits}]{hyperref}
\usepackage[capitalize]{cleveref}

\usepackage{tabularx}
\usepackage{cellspace}
\usepackage{booktabs}
\usepackage{newfloat}
\usepackage{caption}
\usepackage[per-mode=symbol,separate-uncertainty=true]{siunitx}
\usepackage[dvipsnames]{xcolor}
\usepackage{chemformula}
\usepackage{xprintlen}
\usepackage{lipsum}
\usepackage{fancyhdr}
\usepackage{listings}
\usepackage{subcaption}
\usepackage{comment}
\usepackage{soul}

\usetikzlibrary{calc}

\DeclareSIUnit\permille{\text{\textperthousand}}

\crefformat{subequations}{Eqs.~(#2#1#3)}
\Crefformat{subequations}{Eqs.~(#2#1#3)}

\let\originalleft\left
\let\originalright\right
\renewcommand{\left}{\mathopen{}\mathclose\bgroup\originalleft}
\renewcommand{\right}{\aftergroup\egroup\originalright}

\newcolumntype{Y}{>{\centering\arraybackslash}X}
\newcolumntype{d}{c}
\newcolumntype{a}{c}

\fancypagestyle{plain}{
    \lhead{}
    \fancyhead[R]{\thepage}
    \fancyhead[L]{}
    
    \fancyfoot{}
}

\begin{document}
\begin{abstract}
Quantum error correction is needed for quantum computers to be capable of fault-tolerantly executing algorithms using hundreds of logical qubits. Recent experiments have demonstrated subthreshold error rates for state preservation of a single logical qubit. 
In addition, the realization of universal quantum computation requires the implementation of logical entangling gates. 
Lattice surgery offers a practical approach for implementing such gates, particularly in planar quantum processor layouts.
In this work, we demonstrate lattice surgery between two distance-three repetition-code qubits by splitting a single distance-three surface-code qubit. 
Using a quantum circuit fault-tolerant to bit-flip errors, 
we achieve an improvement in the value of the decoded $ZZ$ logical two-qubit observable compared to a similar non-encoded circuit.
By preparing the surface-code qubit in initial states parametrized by a varying polar angle, we evaluate the performance of the lattice surgery operation for non-cardinal states on the logical Bloch sphere and employ logical two-qubit tomography to reconstruct the Pauli transfer matrix of the operation. In this way, we demonstrate the functional building blocks needed for lattice surgery operations on larger-distance codes based on superconducting circuits.
\end{abstract}

\date{\today}

\author{Ilya~Besedin}
\thanks{These authors contributed equally to this work.}
\author{Michael~Kerschbaum}
\thanks{These authors contributed equally to this work.}
\affiliation{Department of Physics, ETH Zurich, 8093 Zurich, Switzerland}
\affiliation{ETH Zurich - PSI Quantum Computing Hub, Paul Scherrer Institute, 5232 Villigen, Switzerland}
\affiliation{Quantum Center, ETH Zurich, 8093 Zurich, Switzerland}

\author{Jonathan~Knoll}
\affiliation{Department of Physics, ETH Zurich, 8093 Zurich, Switzerland}

\author{Ian~Hesner}
\affiliation{Department of Physics, ETH Zurich, 8093 Zurich, Switzerland}
\affiliation{ETH Zurich - PSI Quantum Computing Hub, Paul Scherrer Institute, 5232 Villigen, Switzerland} 
\affiliation{Quantum Center, ETH Zurich, 8093 Zurich, Switzerland}

\author{Lukas~Bödeker}
\author{Luis~Colmenarez}
\affiliation{Institute for Theoretical Nanoelectronics (PGI-2), Forschungszentrum Jülich, 52428 Jülich, Germany}
\affiliation{Institute for Quantum Information, RWTH Aachen University, 52056 Aachen, Germany}

\author{Luca~Hofele}
\author{Nathan~Lacroix}
\author{Christoph~Hellings}
\author{Fran\c{c}ois~Swiadek}
\affiliation{Department of Physics, ETH Zurich, 8093 Zurich, Switzerland}
\affiliation{Quantum Center, ETH Zurich, 8093 Zurich, Switzerland}

\author{Alexander~Flasby}
\author{Mohsen~Bahrami~Panah}
\affiliation{Department of Physics, ETH Zurich, 8093 Zurich, Switzerland}
\affiliation{ETH Zurich - PSI Quantum Computing Hub, Paul Scherrer Institute, 5232 Villigen, Switzerland}
\affiliation{Quantum Center, ETH Zurich, 8093 Zurich, Switzerland}

\author{Dante~Colao~Zanuz}
\affiliation{Department of Physics, ETH Zurich, 8093 Zurich, Switzerland}
\affiliation{Quantum Center, ETH Zurich, 8093 Zurich, Switzerland}

\author{Markus~Müller}
\affiliation{Institute for Theoretical Nanoelectronics (PGI-2), Forschungszentrum Jülich, 52428 Jülich, Germany}
\affiliation{Institute for Quantum Information, RWTH Aachen University, 52056 Aachen, Germany}

\author{Andreas~Wallraff}
\affiliation{Department of Physics, ETH Zurich, 8093 Zurich, Switzerland}
\affiliation{ETH Zurich - PSI Quantum Computing Hub, Paul Scherrer Institute, 5232 Villigen, Switzerland}
\affiliation{Quantum Center, ETH Zurich, 8093 Zurich, Switzerland}

\title{Realizing Lattice Surgery on Two Distance-Three Repetition Codes \\
with Superconducting Qubits}

\maketitle

Scalable quantum computing relies on the ability to correct errors that may occur during computation~\cite{Shor1996}. Topological error correction codes enable the encoding of logical qubits using physical data qubits arranged in a lattice~\cite{Kitaev2003, Dennis2002}, where physical errors in the code are detected using local stabilizer measurements~\cite{Fowler2009, Raussendorf2007a}. In planar codes like the surface code~\cite{Fowler2012}, the stabilizer measurement circuit involves only gates between neighboring qubits, making it well-suited for architectures with 2D local connectivity, such as superconducting qubits.

In such codes, repeated stabilizer measurements allow for the detection and correction of errors, as demonstrated in logical state preservation experiments using surface codes~\cite{Krinner2022, Zhao2022d, Acharya2023} and color codes~\cite{Ryan-Anderson2021, Ryan-Anderson2024, Lacroix2024}. 
Provided the syndrome extraction circuit operates with a sufficiently low error rate, experiments have shown that increasing the size of the qubit lattice, and thereby the code distance, significantly enhances the protection of the logical state~\cite{Bluvstein2023, Acharya2024a}.

Beyond state preservation, an essential component of fault-tolerant quantum computing is the ability to perform operations on logical qubits. This includes single-qubit gates, a subset of which can be straightforwardly realized transversally in some codes like the color code~\cite{Postler2022}, as well as entangling gates between logical qubits. These can also be realized transversally, as demonstrated between distinct distance-three color codes in systems with reconfigurable connectivity, such as trapped ions~\cite{Postler2022, Ryan-Anderson2024} and Rydberg atoms~\cite{Bluvstein2023}. 
However, in systems with fixed local connectivity, this approach introduces significant overhead and cannot be directly applied.

Lattice surgery~\cite{Horsman2012} is a powerful technique which extends topological codes to multiple logical qubits, enabling fault-tolerant gate operations while maintaining a two-dimensional arrangement of physical qubits.
The fundamental operation in lattice surgery involves measuring an observable of two logical qubits, which is executed through merge and split code deformations~\cite{Bombin2009}.
These two operations can be used as building blocks for more complex operations such as logical CNOT gates~\cite{Terhal2015n, Landahl2014} or magic state distillation~\cite{Litinski2019a}.
Lattice surgery primitives have been successfully demonstrated in state teleportation experiments utilizing distance-two surface codes~\cite{Erhard2021}, distance-three color codes~\cite{Ryan-Anderson2024, Lacroix2024}, and in a logical Bell state preparation experiment involving distance-four 3CX and Bacon-Shor codes~\cite{Hetenyi2024}.

Here, we demonstrate  lattice surgery operation on a quantum device consisting of 17 qubits realized with superconducting circuits. 
Specifically, we focus on the split operation applied to a rotated distance-three surface code, which results in an entangled state of two independently operated logical qubits, encoded with bit-flip repetition codes.
The protocol is fault-tolerant for bit-flip but not for phase-flip errors, which remain undetectable after the split. 
By focusing on the boundary region connecting the two logical qubits, we investigate critical aspects of error correction and lattice surgery, such as syndrome correlation analysis during code deformation and logical process tomography, providing insights which can be applied to larger, fully fault-tolerant error-correcting schemes. 
This work highlights the importance of efficient logical characterization of lattice surgery operations as a tool to improve and optimize logical-qubit performance in quantum computing architectures.

\section*{Lattice surgery on a distance-three surface code}

\begin{figure*}[t]
\includegraphics[width=\textwidth]{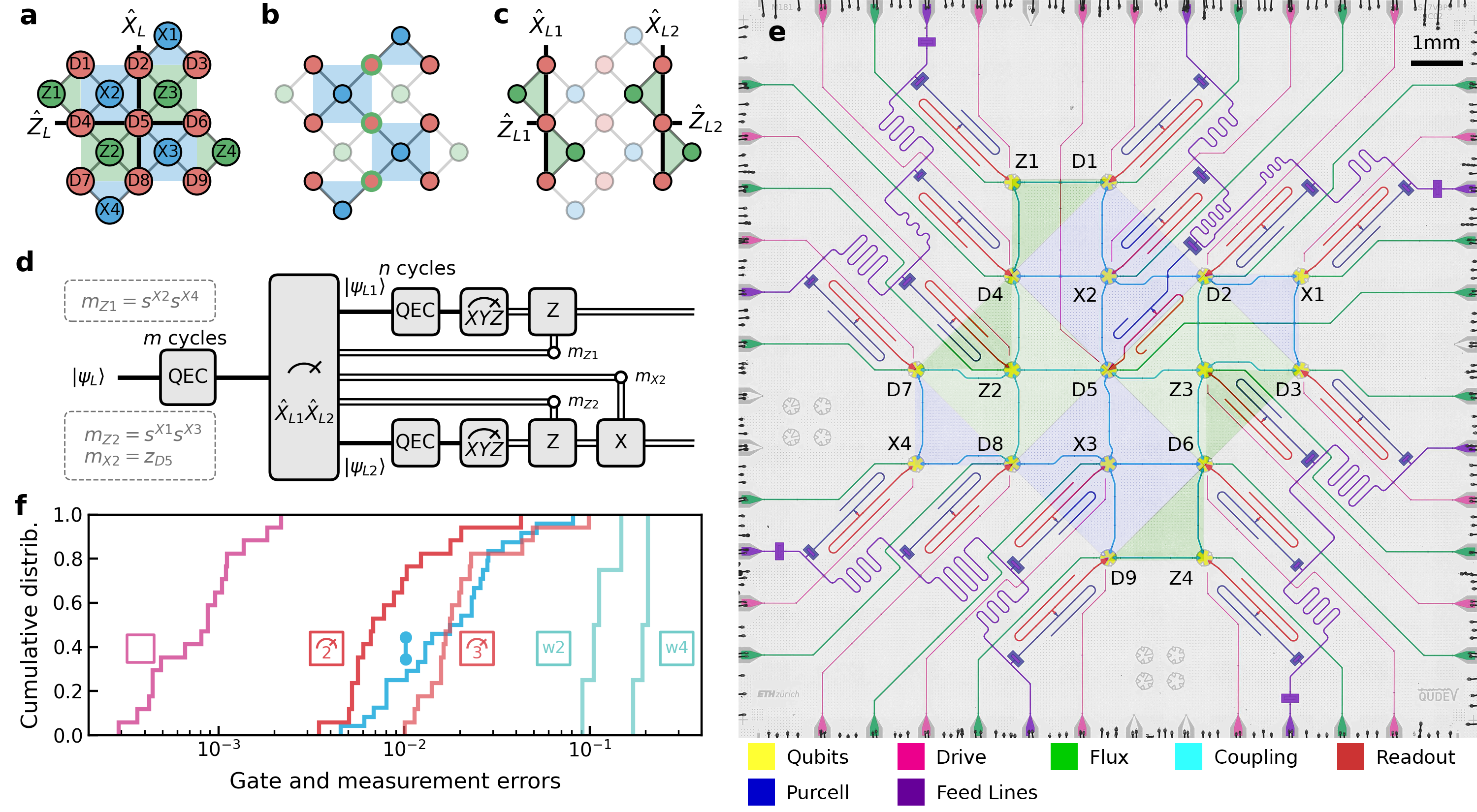}
\centering
\caption{\label{fig:1}Experiment concept, quantum device layout and performance. \textbf{a - c} Conceptual qubit layout and connectivity. The red circles correspond to data qubits, while the green (blue) plaquettes and circles depict $Z$-type ($X$-type) stabilizers and auxiliary qubits associated with them. The logical operator definitions of the distance-three surface code (\textbf{a}) and the two distance-three repetition codes (\textbf{c}) are indicated as solid black lines. Mid-circuit data qubit readout during the lattice-split operation (\textbf{b}) is depicted as green circle outlines. \textbf{d} Schematic experimental sequence and Pauli-frame-update definitions $m_\mathrm{Z1},m_\mathrm{Z2}$ and $m_\mathrm{X2}$. Double lines denote an update of the results of the tomography readouts in post-processing based on measurement outcomes obtained during the lattice split operation. \textbf{e} False-colored optical photograph of the quantum chip. The stabilizer plaquettes are shown with blue and green triangles and squares. \textbf{f} Experimental cumulative distributions of the operation errors of the device: single-qubit gates (pink), two-qubit CZ gates (cyan), readout with two-state (red) and three-state (light red) discrimination and cycle-average syndrome elements for weight-two and weight-four stabilizers (w2 and w4).}
\end{figure*}

In lattice surgery \cite{Bombin2009, Horsman2012}, logical qubits are encoded in patches of distinct data qubits forming a larger underlying lattice.
For the surface code, this lattice is square. 
In each logical qubit, errors are identified through measurements of stabilizers, i.e., products of Pauli-$Z$ or Pauli-$X$ operators acting on data qubits located at the vertices of the lattice plaquettes, with $X$-type and $Z$-type stabilizers arranged in a checkerboard pattern. 
The rotated distance-three surface code implemented in this work consists of a three-by-three grid of data qubits, denoted $\mathrm{D}j$, $j\in\{1,\dots,9\}$, and eight stabilizers. 
Each stabilizer is assigned a unique auxiliary qubit for its measurement. 
The qubit layout and connectivity is shown in Fig.~\ref{fig:1}a.

Error correction is performed by measuring repeated cycles of stabilizers. 
We denote the stabilizer operators as $\hat{S}^{Ai}$, and their measurement outcomes in cycle $N$ as $s^{Ai}_N=\pm1$, where $A\in\{X, Z\}$ indicates the stabilizer type, and $i$ enumerates the stabilizers of one type. 
Syndrome elements are defined as the parity of stabilizer outcomes across two consecutive cycles, where a value of zero indicates even parity, and a value of one indicates odd parity: $\sigma^{Ai}_N = (1-s^{Ai}_N s^{Ai}_{N-1})/2$. 
In the absence of errors, stabilizer measurements remain unchanged between cycles, resulting in syndrome elements of zero. 
If a stabilizer measurement outcome changes between cycles, it signals an error, causing the corresponding syndrome element to flip, i.e., change from 0 to 1, or vice versa, in the subsequent cycle.

Logical operations are performed using code deformations~\cite{Bombin2009}. In our experiment, we perform an $X$-type lattice split on a distance-three surface-code logical qubit. 
We define the logical-qubit $X$ and $Z$ Pauli operators as the products of respective Pauli operators of the middle column $\hat{X}_\mathrm{L} = \hat{X}_\mathrm{D2}\hat{X}_\mathrm{D5}\hat{X}_\mathrm{D8}$ and middle row $\hat{Z}_\mathrm{L} = \hat{Z}_\mathrm{D4}\hat{Z}_\mathrm{D5}\hat{Z}_\mathrm{D6}$ of data qubits, as indicated in Fig.~\ref{fig:1}a. 
After $m+1=4$ cycles of $X$-type stabilizer measurements interleaved with $m=3$ cycles of $Z$-type stabilizer measurements, we read out the middle column of data qubits (D2, D5, D8) in the $Z$ basis, and stop any subsequent $X$-type stabilizer measurements. 
This code deformation step is illustrated in~Fig.~\ref{fig:1}b. The remaining data qubits D1, D3, D4, D6, D7, and D9, and $Z$ stabilizers form two bit-flip repetition codes, as shown in Fig.~\ref{fig:1}c. 
After the split, we execute $n=2$ cycles of $Z$-type stabilizer measurements of the bit-flip codes. 
The chosen values of $m$ and $n$ allow us to isolate the effects of state preparation, split and readout operations. 
We define the logical-qubit operators associated with the bit-flip codes as $\hat{Z}_\mathrm{L1} = \hat{Z}_\mathrm{D4}, \hat{X}_\mathrm{L1} = \hat{X}_\mathrm{D1}\hat{X}_\mathrm{D4}\hat{X}_\mathrm{D7}$, $ \hat{Z}_\mathrm{L2} = \hat{Z}_\mathrm{D6}, \hat{X}_\mathrm{L1} = \hat{X}_\mathrm{D3}\hat{X}_\mathrm{D6}\hat{X}_\mathrm{D9}$. 

As a result of the split operation, the single logical degree of freedom of the distance-three surface
code is transformed into two distinct degrees
of freedom.
The new additional degree of freedom is associated with the $X$-type stabilizers, which are no longer measured, and is used to store logical information. 
Specifically, the product of the four $X$-type stabilizers corresponds to the product of the $\hat{X}$ operators of the remaining data qubits, which equals $\hat{X}_\mathrm{L1}\hat{X}_\mathrm{L2}$. 
For the $\hat{Z}$ operators we can similarly obtain $\hat{Z}_\mathrm{L1}\hat{Z}_\mathrm{L2} = \hat{Z}_\mathrm{L}z_\mathrm{D5}$, where $z_\mathrm{D5}$ is the measurement outcome of the central data qubit. 
Similar to quantum teleportation~\cite{Bennett1993}, the entangled state of the logical qubits depends on the readout outcomes of the auxiliary degrees of freedom used to facilitate the entanglement. 

Throughout this paper, we consider a deterministic version of the split operation.
To that end, we apply a Pauli-frame update to the bit-flip-encoded logical qubits. If the final outcome of $s^{X2}s^{X4}$ ($s^{X1}s^{X3}$) is $-1$, we apply a virtual $Z$ gate to the first (second) logical qubit. Similarly, if $z_\mathrm{D5}=-1$, we apply a virtual $X$ gate to the second logical qubit. The virtual gates are performed by flipping the logical qubit observable outcomes in post-processing.
The experimental sequence, together with the Pauli-frame update, is shown in~Fig.~\ref{fig:1}d. 
Additionally, the syndrome definition must be modified to include the readout outcome of the measured data qubits, see App.~\ref{app:pauliframe} for details. 

For an ideal $X$-type split operation with no error, after the Pauli-frame update, the action on the logical subspace is that of a Hadamard-transformed fanout gate \cite{Hoyer2004}
\begin{equation}
    \label{eq:Fanout}
    \alpha\ket{+} + \beta\ket{{-}} \to \alpha\ket{++} + \beta\ket{{--}}.
\end{equation}

For our experiments, we use a device similar to the one described in Ref.~\onlinecite{Krinner2022}. 
This device consists of 17 flux-tunable transmon qubits with a connectivity designed for implementing a distance-three surface code~\cite{Versluis2017}. 
The qubits are controlled via individual charge lines (shown in pink) for single-qubit gates and flux lines (green) for two-qubit gates. 
The device features readout circuits, consisting of readout-resonator-Purcell-filter-pairs  (red and blue) coupled to shared feedlines (purple) for frequency-multiplexed readout~\cite{Heinsoo2018}, 
see Fig.~\ref{fig:1}e for a false-colored micrograph. The implementations of single-qubit gates, two-qubit gates, and readout are described in Methods Sections~\ref{subsec:sqg}, \ref{subsec:readout} and~\ref{subsec:tqg}. 
We do not reset qubits after their readout, the implications of which are discussed in Refs.~\onlinecite{Krinner2022,Geher2024b}.

Using randomized benchmarking and interleaved randomized benchmarking, we measure average single-qubit and two-qubit gate errors of $(0.09\pm 0.05)\%$ and $(2.2\pm1.7)\%$.
The average two-state and three-state readout assignment fidelity is 98.5\% and 97.5\%, respectively. 
For an integral error metric, we execute a state preservation experiment in the distance-three surface code, obtaining an average syndrome element value of 0.182 for weight-four syndromes and 0.114 for weight-two syndromes (see App.~\ref{app:performance} for details). 
Cumulative distribution functions of the individual gate and readout errors are shown in~Fig.~\ref{fig:1}f together with the average syndrome elements. 

We use a stabilizer measurement circuit that is fault-tolerant against circuit-level noise, described in detail in Methods Section~\ref{subsec:sequence}. Assuming that only one error occurs -- whether during a single-qubit gate, a two-qubit gate, qubit initialization, or readout -- the effect of that error on the logical-qubit observables can be correctly identified using the syndrome elements of the distance-three surface code~\cite{Tomita2014a}.
$Z$-type stabilizer measurements are performed throughout the entire circuit, including preparation of a $Z$-basis cardinal state of the distance-three surface code, the split operation, and the readout of $Z$ observables.
As a result of the fault-tolerant nature of lattice surgery, the $Z_\mathrm{L1}$ and $Z_\mathrm{L2}$ observables are protected from any single error. 
However, for the bit-flip repetition codes, $X$-type stabilizers are not measured, making it impossible to track errors affecting the $X_\mathrm{L1}$ and $X_\mathrm{L2}$ observables after the split operation.

\section*{Logical Bell State Preparation with a Lattice Split}

\begin{figure*}[t!]
\includegraphics[width=\textwidth]{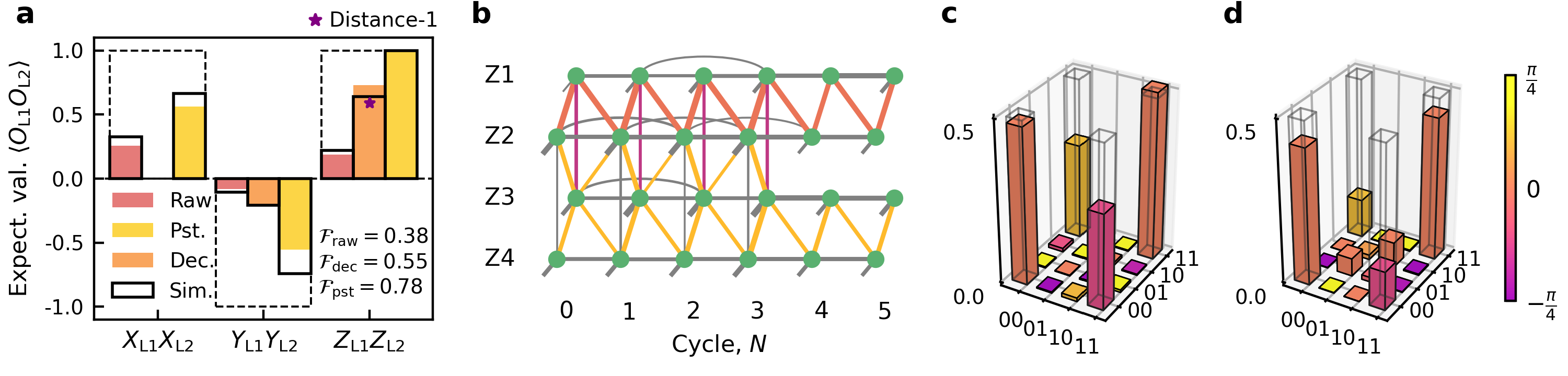}
\centering
\caption{\label{fig:3}Characterization of the lattice-split operation executed on the logical $\ket{0}_\mathrm{L}$ state. 
\textbf{a}~Expectation values of the repetition-code logical-qubit-operator pairs for the split operation. 
Raw (red) and syndrome-postselected values (yellow) are shown for $X_\mathrm{L1}X_\mathrm{L2}, Y_\mathrm{L1}Y_\mathrm{L2}$ and $Z_\mathrm{L1}Z_\mathrm{L2}$-observables, as well as MWPM-decoded values (orange) for $Z_\mathrm{L1}Z_\mathrm{L2}$ and $Y_\mathrm{L1}Y_\mathrm{L2}$. 
We mark ($\star$) an experimentally extracted $Z_\mathrm{L1}Z_\mathrm{L2}$ value of a distance-one experiment for comparison. 
\textbf{b}~Experimentally extracted decoder matching graph for $Z$-syndromes. 
Nodes stand for individual syndrome measurements. 
Each row corresponds to one stabilizer operator in all cycles. 
Edges represent errors that cause the connected syndrome elements to flip. 
Thicker edges denote higher error probability.
Red (orange) edges correspond to errors which flip the $Z_\mathrm{L1}$ ($Z_\mathrm{L2}$) observable, and purple edges correspond to errors which flip both repetition-code observables. Gray edges represent errors that do not affect the logical observables.
See main text for details.
\textbf{c, d}~Density matrix of the logical Bell state extracted from the (c) postselected and (d) decoded logical two-qubit tomography. 
Bar heights show the absolute value of the matrix element, color encodes the phase, and wireframes indicate ideal values.
}
\end{figure*}

We first consider the task of creating a Bell state between two logical qubits. 
In this case, we initialize our logical surface-code qubit in the $\ket{0}_\mathrm{L}$ state.
An ideal lattice-split operation, as described by Eq.~\eqref{eq:Fanout}, yields the following two-qubit observables: $X_\mathrm{L1}X_\mathrm{L2} = +1$, $Z_\mathrm{L1}Z_\mathrm{L2} = +1$, and $Y_\mathrm{L1}Y_\mathrm{L2} = -1$.
Rejecting leakage events (see  Methods Section~\ref{subsec:readout}), we retain $76.8\%$ of the $57\,456$ experimental runs.
We evaluate the logical two-qubit observables as a product of the final data-qubit measurement outcomes.
The measured values deviate significantly from their ideal counterparts stated above (see Fig.~\ref{fig:3}a), due to errors occurring during circuit execution, some of which are detectable during the quantum error correction cycles. 

As a consequence of the bit-flip fault-tolerance of our experiment, we are able to decode $Z$-type syndrome data using a minimum-weight perfect matching (MWPM) decoder~\cite{OBrien2017, Edmonds1965}.
The matching graph is constructed by calculating weights based on correlations between syndrome measurements~\cite{Spitz2018}, accounting for higher-weight error mechanisms.
For this experiment, the matching graph consists of four times six nodes corresponding to the extracted syndromes. 
Edges in the graph represent possible errors.
We observe no statistically significant correlations between syndromes of the first bit-flip code $\sigma^{Z1}_N, \sigma^{Z2}_N$ and the second bit-flip code $\sigma^{Z3}_N, \sigma^{Z4}_N$ after the split operation ($N=3$).
We interpret this as the absence of correlated errors between the two distance-three repetition codes, highlighting the potential for independent state preservation in each logical qubit (Fig.~\ref{fig:3}b).

When we account for bit-flip errors in post-processing, the expectation value of the $Z_\mathrm{L1}Z_\mathrm{L2}$ observable increases from its raw value of $0.189(5)$ to $0.730(3)$ and decreases for $Y_\mathrm{L1}Y_\mathrm{L2}$ from $-0.082(5)$ to $-0.199(5)$. 
While we observe a significant increase in the fault-tolerantly extracted expectation value of $Z_\mathrm{L1}Z_\mathrm{L2}$, the more modest improvement for $Y_\mathrm{L1}Y_\mathrm{L2}$ can be attributed to its susceptibility to non-correctable phase-flip errors (see Fig.~\ref{fig:3}a).
Performing quantum error correction, we improve the Bell state fidelity from $F_\mathrm{raw}=0.382(2)$ to $F_\mathrm{dec}=0.546(2)$.

Furthermore, by postselecting on runs of the experiment where none of the syndrome measurements indicate an error, we operate the experiment in an error detection regime. 
For the fault-tolerantly measured $Z_\mathrm{L1}Z_\mathrm{L2}$ observable, the postselected expectation value improves significantly to $0.998(1)$, approaching the ideal value of $1$. 
In contrast, the $X_\mathrm{L1}X_\mathrm{L2}$ observable shows only a small increase from $0.255(5)$ to $0.56(2)$, and the $Y_\mathrm{L1}Y_\mathrm{L2}$ observable a decrease from $-0.082(5)$ to $-0.55(2)$, again due to phase-flip errors remaining undetectable after the lattice-split operation (see Fig.~\ref{fig:3}a).
The fraction of retained experimental runs after postselection is $6.3\%$ for the $X_\mathrm{L1}X_\mathrm{L2}$ and $Y_\mathrm{L1}Y_\mathrm{L2}$ observables, and $5.5\%$ for the $Z_\mathrm{L1}Z_\mathrm{L2}$ observable. The Bell state fidelity for the postselected dataset is $0.780(6)$, and is mostly limited by the non-fault-tolerantly extracted two-qubit observables.

To highlight the effectiveness of our bit-flip error correction scheme, we perform a non-encoded variant of the split experiment using three data qubits, two $X$-type auxiliary qubits, and $m+1=4$ cycles of $X$-syndrome extraction, similar to the distance-three implementation. 
Despite the increased circuit complexity of the error-corrected code it shows an average $\hat{Z}_\mathrm{L1}\hat{Z}_\mathrm{L2}$ expectation value of $0.730(3)$, which is significantly closer to the ideal value of unity than $0.591(8)$ for the non-error-corrected variant, see star in Fig.~\ref{fig:3}a.
Details of the non-error-corrected implementation can be found in App.~\ref{app:distance1}. 

We compare the measured expectation values to simulations of the experimental gate sequence using a Pauli error model based on the independently measured
gate errors and qubit coherence times.
Using stabilizer circuit simulations~\cite{Gidney2021}, we sample measurement outcomes of the circuit with the applied error model.
The simulated two-qubit observables, indicated in Fig.~\ref{fig:3}a with wireframes, show close agreement with their measured counterparts. 
We find an average norm-based fidelity~\cite{Liang2019} between the experimental and simulated logical density matrices of $\mathcal{F}_2 = 96\%$ and discuss details of the simulation in App.~\ref{app:Simulation}. 

To characterize the state of the logical qubits after the split operation, we perform logical state tomography~\cite{Postler2022, Gupta2024}. 
With the logical Bell state prepared as described above, we measure each combination of the logical two-qubit Pauli observables, covering nine measurement bases. 
For this measurement, we also correct errors that affect the $I_{L1}X_{L2}$, $X_{L1}I_{L2}$ observables. The decoding is described in App.~\ref{app:decoding}.
We reconstruct the logical two-qubit density matrix for the postselected as well as the decoded dataset. 
For both datasets, we find that the entries on the diagonal of the density matrix remain close to their ideal values of $0.5$ and $0$ for the $\ket{00}_\mathrm{L}$, $\ket{11}_\mathrm{L}$ and $\ket{01}_\mathrm{L}$, $\ket{10}_\mathrm{L}$ states, respectively, which is a consequence of the protocol's fault-tolerance with respect to bit-flip errors.
However, the coherences, i.e. off-diagonal elements of the density matrix, are degraded due to phase-flip errors remaining undetectable after the lattice split, see Fig.~\ref{fig:3}c~and~d for the postselected and decoded datasets, respectively. 
Additionally, we observe a phase rotation in one of the logical qubits, indicated by a phase shift of $0.11\pi$ in the off-diagonal entries compared to the ideal value of $0$. 
This phase error is discussed in more detail in App.~\ref{app:PhaseRotation}.
\section*{Splitting Arbitrary Logical States}

\begin{figure}[t!]
\includegraphics[width=0.5\textwidth]{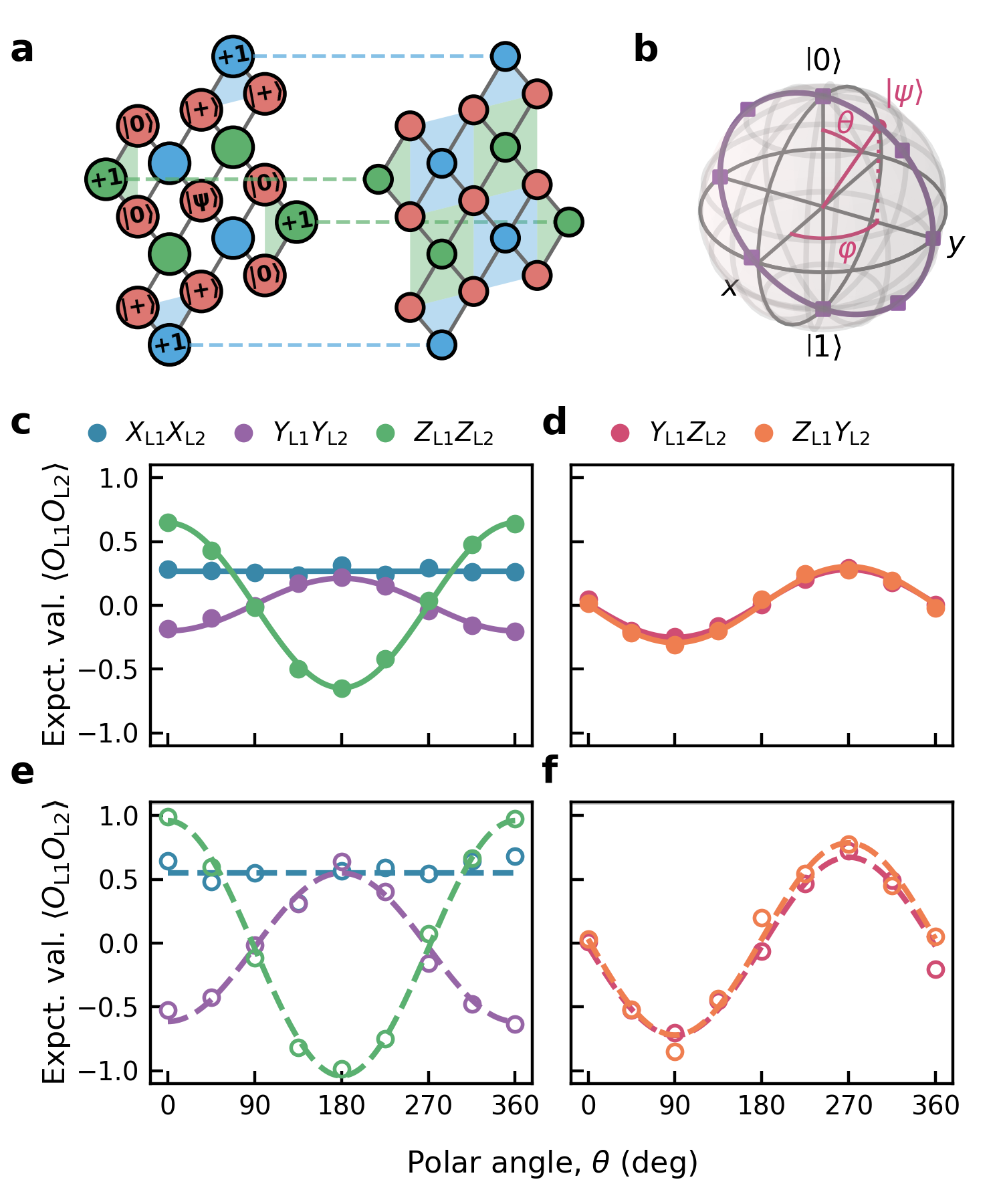}
\centering
\caption{Lattice split of arbitrary states. \textbf{a} Illustration of the arbitrary state preparation sequence for the distance-three surface code (see text for details). 
\textbf{b} Parametrization of the arbitrary state in the $yz$-plane by the polar angle $\theta$. 
Filled purple squares in the $yz$-plane indicate state preparations used as input states for the lattice-split operation. 
Non-zero two-qubit observables versus polar angle $\theta$ in the $yz$ plane for \textbf{c, d} decoded and \textbf{e, f} postselected observables. 
The straight blue solid and dashed lines represent the average expectation value of $X_\mathrm{L1}X_\mathrm{L2}$, while the other solid and dashed lines correspond to sinusoidal fits of the $Y_\mathrm{L1}Y_\mathrm{L2}$, $Z_\mathrm{L1}Z_\mathrm{L2}$, $Z_\mathrm{L1}Y_\mathrm{L2}$, and $Y_\mathrm{L1}Z_\mathrm{L2}$ observable expectation values.}
\label{fig:4}
\end{figure}

The lattice split is a quantum operation acting on an encoded qubit. 
To characterize its action on different initial logical states, we prepare arbitrary input states of the distance-three surface-code logical qubit via state injection~\cite{Lao2022, Ye2023a}. 
We initialize the central data qubit (D5) in the target state $\psi$, and the other data qubits in either the $\ket{+}$ or $\ket{0}$ states, such that the weight-two stabilizers of the surface code are well-defined. 
In Fig.~\ref{fig:4}a the initial states of the data qubits are shown with the labels in the red circles. 
Stabilizers which are well-defined are indicated by colored plaquettes (blue or green). 
The labels in the blue and green circles correspond to the initial values of the stabilizers for this particular initialization of the data qubits. 
After the first error correction cycle, all stabilizers are in a well-defined state, indicated by filled plaquettes in Fig.~\ref{fig:4}a. 
The dashed lines between stabilizers indicate valid syndromes which can be used to detect errors. 

While this procedure is not fully fault-tolerant, it allows us to postselect on runs where no errors are detected by the weight-two stabilizers after the first stabilizer measurement cycle. 
Retaining 48\% of the data on average,  we prepare arbitrary logical states with an average postselected fidelity of $97.0\%$ (see App.~\ref{app:arbstate} for details). 

We evaluate the action of the lattice-split operation on arbitrary logical states in the $yz$-plane of the Bloch sphere, 
parameterized by the polar angle $\theta$ with a fixed value of the azimuthal angle $\varphi=90^\circ$ (Fig.~\ref{fig:4}b). 
We characterize the final state of the bit-flip codes using logical state tomography.
Experimental runs with errors detected in the first round of stabilizer measurements are discarded.

\begin{figure*}[t!]
\includegraphics[width=\textwidth]{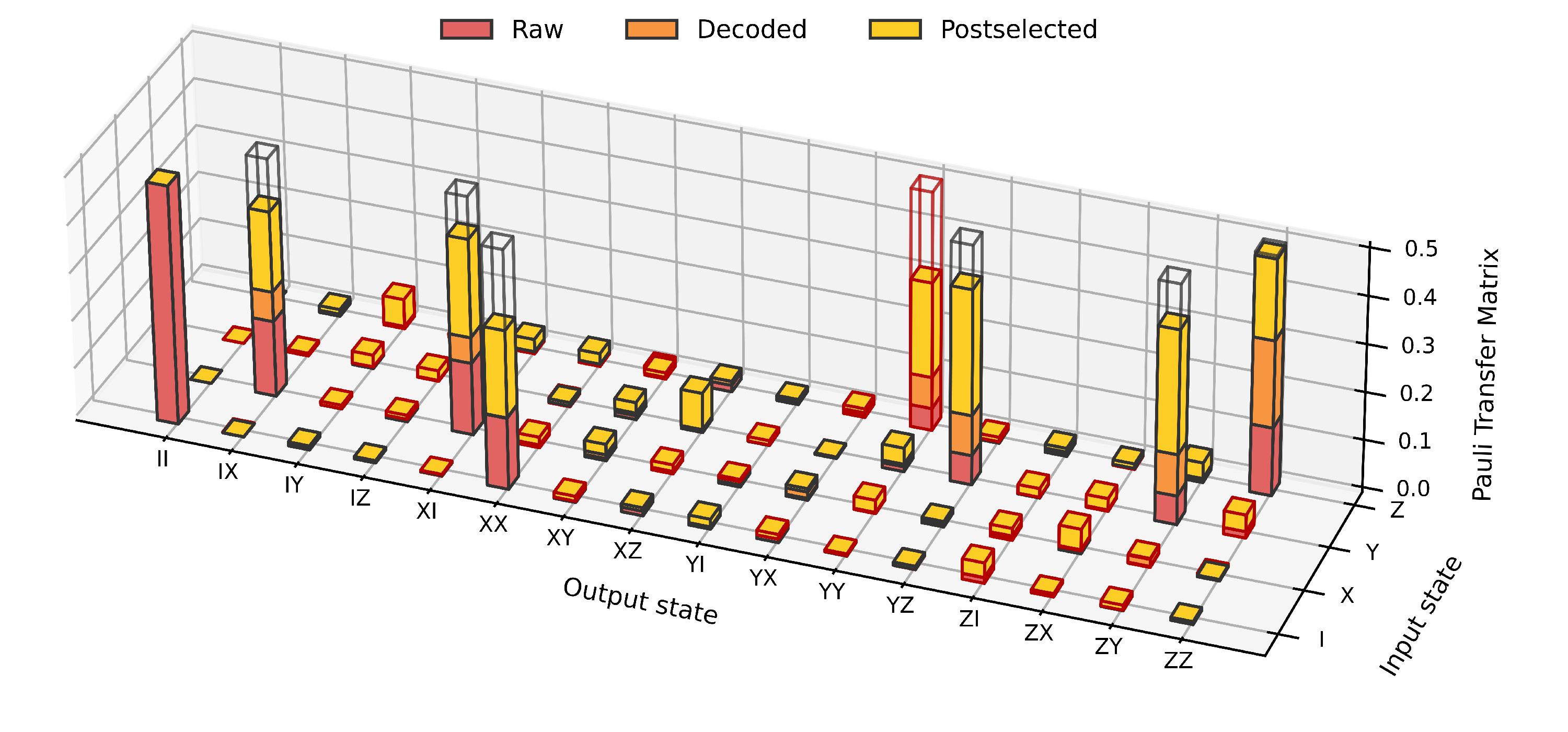}
\centering
\caption{\label{fig:5}Pauli transfer matrix (PTM) of the split operation reconstructed from raw (red), decoded (orange) and postselected (yellow) logical observable outcomes.
The input state operators refer to the distance-three surface-code observable $O_\mathrm{L}$, and the output state operators refer to joint observables of the two bit-flip repetition codes $O_\mathrm{L1}O_\mathrm{L2}$.
The height of each colored bar or wireframe indicates the absolute value of the corresponding PTM element. 
Wireframe color indicates the sign of the PTM entry (black for positive, red for negative). 
The empty wireframes show the PTM entries expected for an ideal split operation.}
\end{figure*}

For the polar angle $\theta=0$, the initial state is $\ket{0}_\mathrm{L}$, and the outcomes are in line with the logical Bell state.
As a function of polar angle $\theta$, the $X_\mathrm{L1}X_\mathrm{L2}$ observable remains close to constant, independent of the initially prepared state, 
while the logical $Z_\mathrm{L1}Z_\mathrm{L2}$, $Y_\mathrm{L1}Y_\mathrm{L2}$, $Y_\mathrm{L1}Z_\mathrm{L2}$, and $Z_\mathrm{L1}Y_\mathrm{L2}$ observables show sinusoidal dependencies on $\theta$, as expected from Eq.~\eqref{eq:Fanout}. We observe the same qualitative features when performing either error correction or postselecting on runs where no errors were detected. However, for postselection the contrast is higher for all observables.
The highest contrast of 0.646(5) for the error-corrected observable and 0.97(2) for the postselected observable is obtained for $Z_\mathrm{L1}Z_\mathrm{L2}$, consistent with the fault-tolerant nature of the protocol. 
The $X_\mathrm{L1}X_\mathrm{L2}$ observable expectation values are independent of $\theta$, showing an average value of 0.267(9) for the error-corrected and 0.55(3) for the postselected outcomes. The contrasts for $Y_\mathrm{L1}Y_\mathrm{L2}$, $Y_\mathrm{L1}Z_\mathrm{L2}$, and $Z_\mathrm{L1}Y_\mathrm{L2}$ also show a reduced magnitude.
The expectation values of the indicated two-qubit observables evaluated with error correction are shown in Fig.~\ref{fig:4}c~and~d, and with postselection on no detected errors in Fig.~\ref{fig:4}e~and~f. 

To further characterize the split operation acting on the logical qubit, we perform logical quantum process tomography~\cite{Postler2022, Gupta2024}. 
By preparing the distance-three surface code in eigenstates of the $\hat{X}_\mathrm{L}$, $\hat{Y}_\mathrm{L}$ and $\hat{Z}_\mathrm{L}$ operators using the approach described above, applying the lattice-split operation, and performing a tomographic readout with nine combinations of logical single-qubit observables, we reconstruct the process map.

The transformation of a single distance-three surface-code logical qubit into two distance-three bit-flip repetition-code logical qubits after the split is described by a $4 \times 16$ process matrix. 
For the reconstruction of the non-square process matrix, we follow the conventions described in Ref.~\onlinecite{Schwartz2016}.
We reconstruct the Choi matrix with the positive semi-definite constraint, ensuring complete positivity of the process map. 
Reconstruction is performed for three sets of observables: raw, decoded and postselected on no detected errors in the entire run.
The resulting process fidelity for the raw outcome is 0.310(12). Decoding improves the process fidelity to 0.442(12). Postselection yields a process fidelity of 0.78(3), while retaining only 3.3\% of the experimental runs.
In Fig.~\ref{fig:5}, we show the Pauli transfer matrix representations of the raw (red), decoded (orange), and postselected (yellow) process maps, with black (red) wireframes around entries indicating positive (negative) matrix values.
Furthermore, in the analysis, we correct for the coherent phase rotation of $0.11\pi$ observed in the first bit-flip repetition-code qubit (see App.~\ref{app:PhaseRotation} for details).
The eight entries that are expected to take non-zero values are highlighted by empty wireframes with the ideal absolute value of 0.5. 
We find good qualitative agreement between the reconstructed Pauli transfer matrix and the results of our stabilizer simulation, see App.~\ref{app:Simulation} for details.

In this work, we have experimentally demonstrated a lattice-split operation on a distance-three surface code, exploring lattice surgery for entangling logical qubits.
Additionally, our experiments highlight the power of logical process tomography in identifying and mitigating effective errors at the logical level.
We anticipate that future experiments implementing lattice surgery protocols with two distance-three (or larger) surface codes will achieve fault tolerance not only for bit-flip errors but also for phase-flip errors, further reducing both coherent and incoherent error rates. 
The characterization techniques explored in this work will be valuable for future realizations of lattice surgery.

\section*{Acknowledgements}
We are grateful for valuable discussions with S. Krinner on conceptual aspects of this work. Research was sponsored by IARPA and the Army Research Office, under the Entangled Logical Qubits program, and was accomplished under Cooperative Agreement Number W911NF-23-2-0212, by the Swiss State Secretariat for Education, Research and Innovation (SERI) under contract number UeM019-11, by Innosuisse via the Innovation project (104.020 IP-ICT / Agreement Nr. 2155012229), by the SNSF R'equip grant 206021-170731, by the Baugarten Foundation and the ETH Zurich Foundation, and by ETH Zurich. 
The views and conclusions contained in this document are those of the authors and should not be interpreted as representing the official policies, either expressed or implied, of IARPA, the Army Research Office, or the U.S. Government. 
The U.S. Government is authorized to reproduce and distribute reprints for Government purposes notwithstanding any copyright notation herein.
L.B., L.C., and M.M. additionally acknowledge support from the German Research Foundation (DFG) under Germany's Excellence Strategy ``Cluster of Excellence Matter and Light for Quantum Computing (ML4Q) EXC 2004/1'' 390534769, the ERC Starting Grant QNets through Grant No. 804247, and the Munich Quantum Valley (K-8), which is supported by the Bavarian state government with funds from the Hightech Agenda Bayern Plus.

\section*{Author Contribution}
I.B. and M.K. planned and performed the experiments, and analyzed the data. 
F.S. designed the device, and A.F., M.B.P. and D.C.Z. fabricated the device. 
I.B., M.K., J.K, L.H., N.L. and C.H. developed the experimental software framework. 
M.K. designed and built elements of the experimental setup with contributions from I.B. 
I.B., M.K., and L.H. characterized and calibrated the device and the experimental setup. 
I.H. and I.B. performed the numerical simulations with contributions from L.B, L.C., and M.K and guided by M.M. 
I.B., M.K., J.K., and I.H. prepared the figures for the manuscript, and 
I.B., M.K., I.H., and A.W. wrote the manuscript with inputs from all co-authors. 
M.M. and A.W. supervised the work.

\section*{Competing Interests}
The authors declare no competing interests.
\section*{Methods}
\label{app:methods}

\subsection{Idling and Single-Qubit Gates}
\label{subsec:sqg}

During idling and single-qubit gate operations, we nominally apply an integer flux in units of the flux quantum to the asymmetric superconducting quantum interference devices (SQUIDs) of the auxiliary qubits, and half-integer flux to the SQUIDs of the data qubits. 
This biases the qubits at the upper and lower first-order flux-insensitive frequencies, respectively.
At these bias points, we measure energy relaxation times $T_1$ between $24$ and $78\,\mu$s and coherence times $T_2^E$ ranging from $12.4$ to $138.9\,\mu$s. 
Due to a two-level system defect at the upper flux-insensitive point of X3, this qubit was instead operated at its lower flux-insensitive point. 
For single-qubit gates, we apply $48\,$ns derivative removal by adiabatic gate (DRAG) microwave pulses~\cite{Motzoi2009, Lazar2023} via a coplanar wave\-guide capacitively coupled to the qubit island. 
We compensate for microwave crosstalk as described in Ref.~\onlinecite{Krinner2022} for qubit pairs where the crosstalk has been shown to adversely affect gate fidelity.

\subsection{Qutrit Readout and Leakage Rejection}
\label{subsec:readout}

\begin{figure*}[ht!]
\includegraphics[width=1\textwidth]{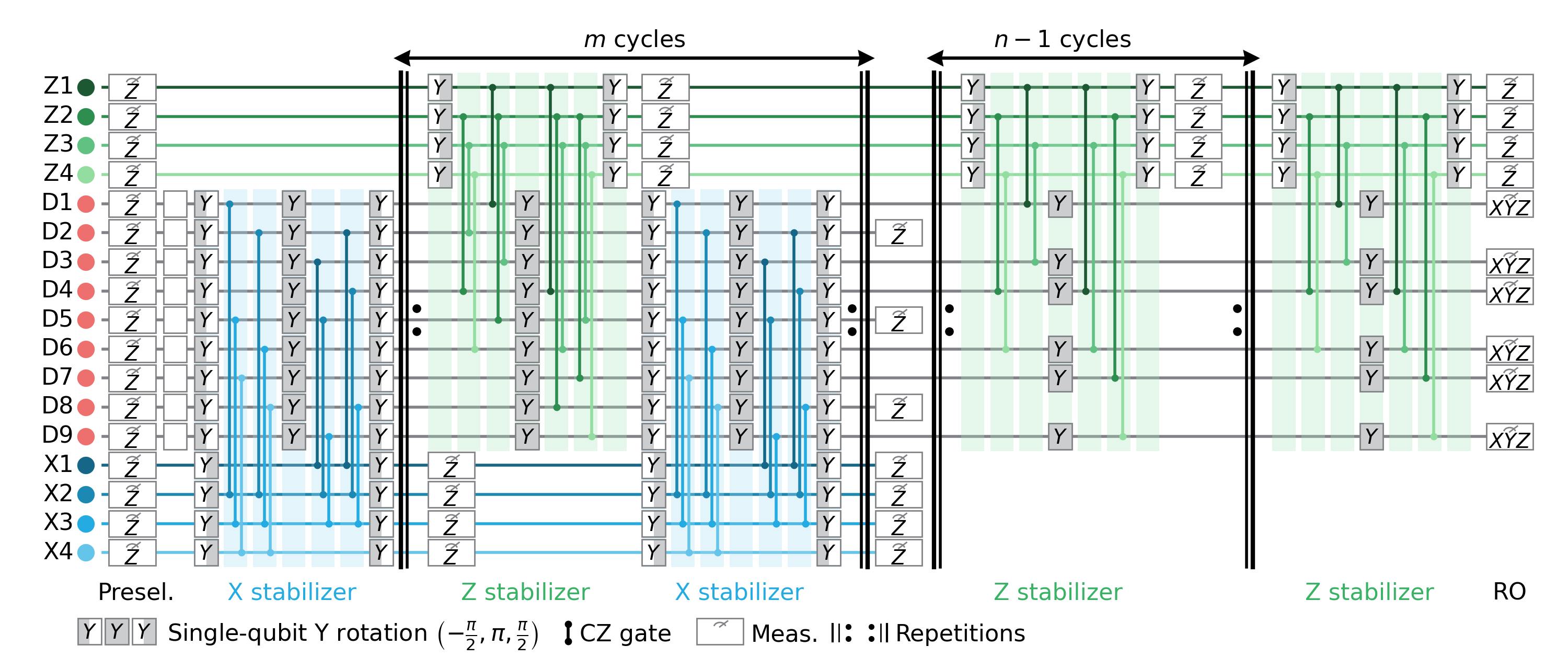}
\centering
\caption{\label{fig:gate_sequence}Experimental sequence of the lattice-split operation.
Measurements are performed in the $Z$ basis, except for the final logical quantum
state tomography performed on the data qubits constituting the two repetition-code qubits, and indicated by $XYZ$. Empty boxes stand for single-qubit gates during state preparation. 
}
\end{figure*}

We measure the qubit states using dispersive single-shot readout via a capacitively coupled readout circuit~\cite{Walter2017}. 
The readout is executed with Gaussian-filtered square microwave pulses applied to the feedline. The pulse frequency is optimized to maximize the separation of the complex scattering amplitudes of the readout signals in the IQ-plane for the prepared transmon states. 
After choosing the readout frequency, the amplitude and duration of the pulse are tuned for maximum single-shot assignment fidelity, resulting in durations ranging from $175$~to~$300\,$ns across the 17 qubits.
By measuring the response of the transmitted readout signal in the time domain, with the qubit prepared in its ground, first, and second excited states, we extract optimal weights for the qubit readout~\cite{Gambetta2007}. 
Additionally, for qubit X3, we use flux-pulse-assisted readout~\cite{Swiadek2024}, reducing the qubit-readout-resonator detuning, and increasing the effective coupling of the readout resonator mode to the readout feedline for improved performance. 

In our current implementation, we do not correct for leakage errors.
We discard all runs of the experiment where a qubit is detected in a leaked state.
In future experiments, leakage reduction units~\cite{McEwen2021a, Lacroix2023} can be used to convert leakage errors into Pauli errors, which has been shown to improve error correction performance for codes with distance larger than three~\cite{Acharya2024a}. 

\subsection{Two-Qubit Gates}
\label{subsec:tqg}

We implement two-qubit gates by leveraging the coherent evolution of neighboring qubits between the $\ket{11}$ and $\ket{02}$ states~\cite{DiCarlo2009}. 
This is achieved using a static capacitive coupling mediated by a coplanar waveguide resonator, with fast flux pulses applied to the SQUIDs of both qubits to bring the states into resonance at a desired interaction frequency within their tunability range. 
To mitigate distortions of the control signals, we use infinite impulse response (IIR) and finite impulse response (FIR) filters~\cite{Rol2020}, along with flux crosstalk compensation as described in Ref.~\onlinecite{Krinner2022}. 
We operate the gate using net-zero pulses~\cite{Rol2019}, with the duration between the two halves of the pulses fine-tuned to adjust the conditional phase~\cite{Negirneac2021}. 
Using defect-mode swap spectroscopy~\cite{Lisenfeld2015}, we scan for the presence of two-level defects and select interaction frequencies that minimize energy loss to these defects while maintaining sufficient spectral distance from transition frequencies of neighboring qubits to avoid spectator errors~\cite{Krinner2020}. 
The average duration of the CZ gate is $101.5\,$ns. 
Additional $20\,$ns buffers are added at the beginning and the end of the net-zero pulse.

\subsection{Gate Sequence of Lattice Split Operation}\label{subsec:sequence}

The stabilizer measurements are realized using a gate sequence similar to the one  in Ref.~\onlinecite{Krinner2022}, consisting of single-qubit $\pi/2$ and $\pi$ rotations, CZ gates, and mid-circuit auxiliary-qubit measurements. 
Within each cycle, active stabilizers of the same type are measured simultaneously. 
The gate sequence for a stabilizer measurement consists of eight time steps, in which all gates within the same step are executed in parallel.
The first and seventh steps are $\pi/2$ rotations on the auxiliary qubits. 
For $X$-type stabilizers, $\pi/2$ pulses are also applied to the data qubits.
The second, third, fifth and sixth steps contain two-qubit gates. The order of the two-qubit gates is chosen to maintain fault tolerance~\cite{Tomita2014a}. Gates involving inactive qubits, that is, $X$-type auxiliary qubits or data qubits in the middle row after the split, are not executed.
We incorporate dynamical decoupling by applying echo pulses to the data qubits in the fourth step, mitigating low-frequency noise on the flux control lines and spectator errors~\cite{Krinner2020}. 
Finally, the eighth step is the auxiliary-qubit readout. 
During surface-code cycles, we employ a pipelined gate scheduling~\cite{Versluis2017}: a new stabilizer measurement begins before the readout step of the other stabilizer type completes.
The duration between the start of successive stabilizer measurement cycles of the same type is $1.66~\mu\mathrm{s}$.
Data qubit readout, both mid-circuit during the code deformation, and the final measurement, is performed in parallel with the auxiliary-qubit readout.
The gate sequence for the lattice-split experiment includes a preselection readout, state preparation, repeated stabilizer measurements, and final data-qubit readout (Fig.~\ref{fig:gate_sequence}). 
The preselection readout is used to discard runs in which any qubit is detected in the excited or leaked state at the beginning of the sequence.
No state preparation pulses are required for the logical $\ket{0}_\mathrm{L}$ state; for the arbitrary state split, single-qubit gates are executed according to the states shown in Fig.~\ref{fig:4}b.

\section*{Supplementary Information}
\appendix
\section{Pauli Frame Update}
\label{app:pauliframe}

During a code deformation, data qubits can be added to or removed from stabilizers. 
For a code deformation to be fault-tolerant, any data qubit added to an existing $X$-type ($Z$-type) stabilizer must be initialized in an eigenstate of the $\hat{X}$ ($\hat{Z}$) operator \cite{Horsman2012}. 
In the absence of errors, the stabilizer measurement outcome after adding the new data qubit can be predicted from the previous cycle's outcome and the chosen initial state of the new qubit.
Conversely, removing a data qubit from a stabilizer while acquiring a syndrome element from that stabilizer requires the data qubit to be read out in the stabilizer basis.
Depending on the initialization state or the readout outcome, the code deformation can result in a flip of the stabilizer. 
For our split experiment, the code deformation involves reading out three data qubits. 
The stabilizers $\hat{S}^{\mathrm{Z}2}$ and $\hat{S}^{\mathrm{Z}3}$ are defined both before and after the code deformation, but their data-qubit composition changes during the $X$-type split.
To accommodate the changing stabilizer compositions, we use a modified definition of syndrome elements in error correction cycle $N=3$ by multiplying the stabilizer values with the mid-circuit data-qubit readout outcomes $z_\mathrm{D2}$, $z_\mathrm{D5}$, and $z_\mathrm{D8}$:
\begin{equation}
\begin{aligned}
    \sigma_4^\mathrm{Z2} = (1-s^\mathrm{Z2}_4s^\mathrm{Z2}_3z_\mathrm{D5}z_\mathrm{D8})/2, \\
    \sigma_4^\mathrm{Z3} = (1-s^\mathrm{Z3}_4s^\mathrm{Z3}_3z_\mathrm{D2}z_\mathrm{D5})/2.
\end{aligned}
\end{equation}

To obtain deterministic logical observable outcomes, we apply a Pauli-frame update to the logical-qubit operators. As a post-processing operation, the Pauli-frame update, shown in Fig.~\ref{fig:1}d, can be expressed as
\begin{align}
\label{eq:PauliFrameUpdate}
\begin{split}
\hat{X}_\mathrm{L1} &\to \hat{X}_\mathrm{L1}s^\mathrm{X2}_4 s^\mathrm{X4}_4, \\
\hat{X}_\mathrm{L2} &\to \hat{X}_\mathrm{L2}s^\mathrm{X1}_4 s^\mathrm{X3}_4, \\
\hat{Z}_\mathrm{L1} &\to \hat{Z}_\mathrm{L1}, \\
\hat{Z}_\mathrm{L2} &\to \hat{Z}_\mathrm{L2}z_\mathrm{D5}.
\end{split}
\end{align}

The transformation involves the $X$-type stabilizer outcomes, which are measured $m+1=4$ times before the split. In the absence of errors, the syndrome elements of the same stabilizer are identical across cycles. However, errors occurring before the last cycle will flip the stabilizer measurement outcomes along with the logical observables. To obtain the best estimate of the final stabilizer outcome we use $s^\mathrm{Xi}_4$ for the Pauli-frame update.

\begin{figure}[t!]
\includegraphics[width=0.5\textwidth]{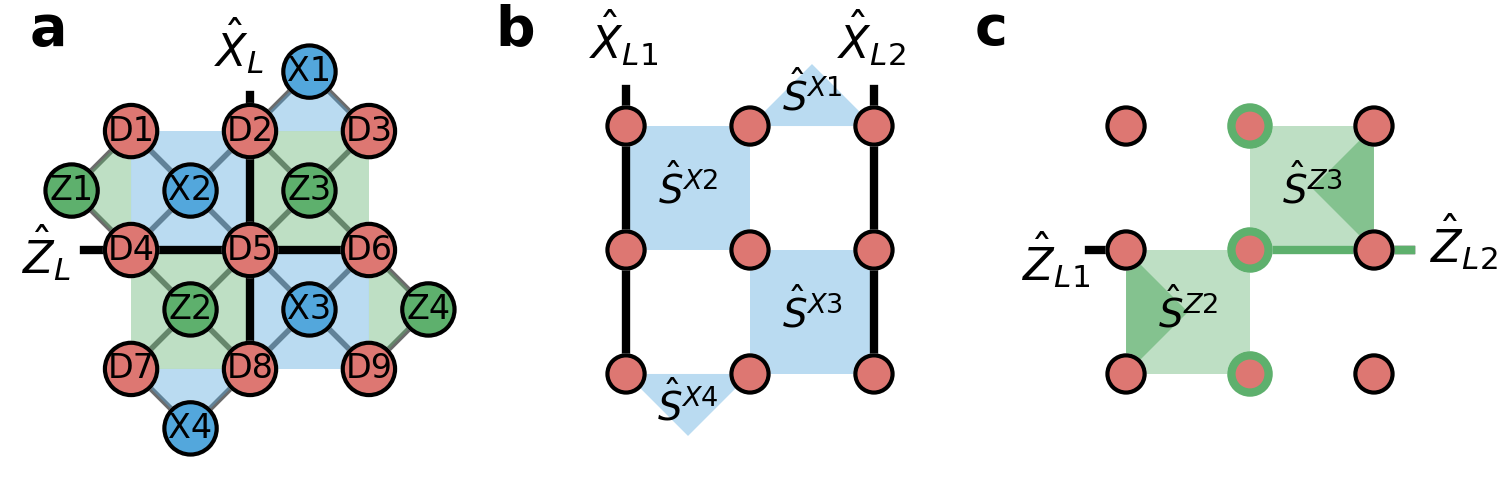}
\centering
\caption{\label{fig:sb1}Definitions of logical-qubit operators of the distance-three surface code and the bit-flip repetition codes and their relation to data-qubit readout and stabilizers. 
The red circles indicate data qubits, green circles $Z$-type and blue circles $X$-type auxiliary qubits. 
\textbf{a}~Naming scheme of data and auxiliary qubits, as well as the definition of the $\hat{Z}_\mathrm{L}$ and $\hat{X}_\mathrm{L}$ for the distance-three surface code.
\textbf{b}~Definition of $X$-type stabilizers and repetition-code $X$-type observables. 
Black vertical lines show strings of data qubits in the definitions of $\hat{X}_\mathrm{L1}$, and $\hat{X}_\mathrm{L2}$. 
For each $X$-type stabilizer, a light-blue rectangle illustrates which data qubits' parity is included in the stabilizer.
\textbf{c}~Definition of repetition-code $Z$-type logical observables. 
The black lines show the data qubits, whose $\hat{Z}$ operators are in the definition of the $\hat{Z}_\mathrm{L1}$ and $\hat{Z}_\mathrm{L2}$ logical-qubit operators.
The green line extending from $\hat{Z}_\mathrm{L2}$ shows that the central data-qubit readout outcome is included in the Pauli-frame update for $\hat{Z}_\mathrm{L2}$. The light-green and dark-green plaquettes indicate the modified $Z$-type stabilizers before and after the split operation, respectively.}
\end{figure}

To motivate the modification of the logical-qubit definitions given by~\cref{eq:PauliFrameUpdate}, we consider the definitions of the $X$-type logical operators $\hat{X}_\mathrm{L}$, $\hat{X}_\mathrm{L1}$, and $\hat{X}_\mathrm{L2}$. 
For a surface code, these definitions are equivalent up to a product of stabilizer operators (Fig.~\ref{fig:sb1}a and b). Specifically, 
\begin{equation}
\begin{aligned}
    \hat{X}_\mathrm{L}\hat{X}_\mathrm{L1} = \hat{S}^\mathrm{X2}\hat{S}^\mathrm{X4}, \\
    \hat{X}_\mathrm{L}\hat{X}_\mathrm{L2} = \hat{S}^\mathrm{X1}\hat{S}^\mathrm{X3}.   
\end{aligned}
\end{equation}
These relations hold until the middle column of data qubits is read out in the $Z$ basis, after which the definitions of $\hat{X}_\mathrm{L}$ and $\hat{S}^{\mathrm{X}i}$ no longer refer to quantities that can be evaluated. By substituting the operators $\hat{S}^{\mathrm{X}i}$ with their final measured outcomes $s^{\mathrm{X}i}_4$, we obtain the Pauli-frame updates for the $X$ observables in~\eqref{eq:PauliFrameUpdate}.

For the $Z$-type logical observables, shown in Fig.~\ref{fig:sb1}c, we have 
\begin{equation}
\hat{Z}_\mathrm{L} = \hat{Z}_\mathrm{D4}\hat{Z}_\mathrm{D5}\hat{Z}_\mathrm{D6}.
\end{equation}
After reading out D5 in the $Z$ basis, the operator $\hat{Z}_\mathrm{D5}$ can be replaced by the measurement outcome $z_\mathrm{D5}$. 
To ensure the deterministic relation $\hat{Z}_\mathrm{L} = \hat{Z}_\mathrm{L1}\hat{Z}_\mathrm{L2}$, the readout outcome of D5 must be included in the definition of either logical operator $\hat{Z}_\mathrm{L1}$ or $\hat{Z}_\mathrm{L2}$. 
We choose to append it to $\hat{Z}_\mathrm{L2}$.
The final expressions for the Pauli-frame-updated $Z$ observables of the bit-flip codes are
\begin{equation}
\begin{aligned}
\hat{Z}_\mathrm{L1} = \hat{Z}_\mathrm{D4}, \\
\hat{Z}_\mathrm{L2} = \hat{Z}_\mathrm{D6}z_\mathrm{D5}. \\
\end{aligned}
\end{equation}

Due to the appended measurement outcome $z_\mathrm{D5}$, the errors corresponding to edges between Z2 and Z3 in the decoder graph in~Fig.~\ref{fig:3}d flip the $Z_\mathrm{L2}$ observable.

\section{Performance of Logical Qubit Idling}
\label{app:performance}

We characterize the performance of our distance-three surface-code qubit using the state preservation experiment of Ref.~\onlinecite{Krinner2022}. 
As an integral characteristic of the physical error rate, we determine the average syndrome values across $Z$-type and $X$-type stabilizers~\cite{Hesner2024}. 
For the $Z$-type stabilizers, we prepare the data qubits in the $\ket{0}^{\otimes 9}$ and $\hat{X}_\mathrm{L}\ket{0}^{\otimes 9}$ states. 
For the $X$-type stabilizers, we prepare the data qubits in the $\ket{+}^{\otimes 9}$ and $\hat{Z}_\mathrm{L}\ket{{+}}^{\otimes 9}$ states. The cycle duration is $1.66~\mu\mathrm{s}$.

After initializing the data qubits, we perform 20 cycles of interleaved $X$-type and $Z$-type stabilizer measurements. 
The gate sequence is similar to the first half of the one used for the lattice-split experiment, shown in Fig.~\ref{fig:gate_sequence}, but omits the final round of $X$-type stabilizer measurements.
Finally, we read out the data qubits in the same basis as the initial state preparation.

The initial values of the stabilizers are defined by the initialized data-qubit states, while the final values of the stabilizers are determined from the data-qubit readout outcomes. 
The first syndrome elements are calculated as parities of the stabilizer values determined by state preparation and the outcome of the first cycle of stabilizer measurements. 
The final syndrome element is calculated as the parity of the data-qubit readout outcomes and the outcome of the last cycle of stabilizer measurements. 
All other syndrome elements, also referred to as ``bulk'' syndromes, are calculated as the parities of subsequent stabilizer measurement outcomes. 

\begin{figure}[t!]
\includegraphics[width=0.5\textwidth]{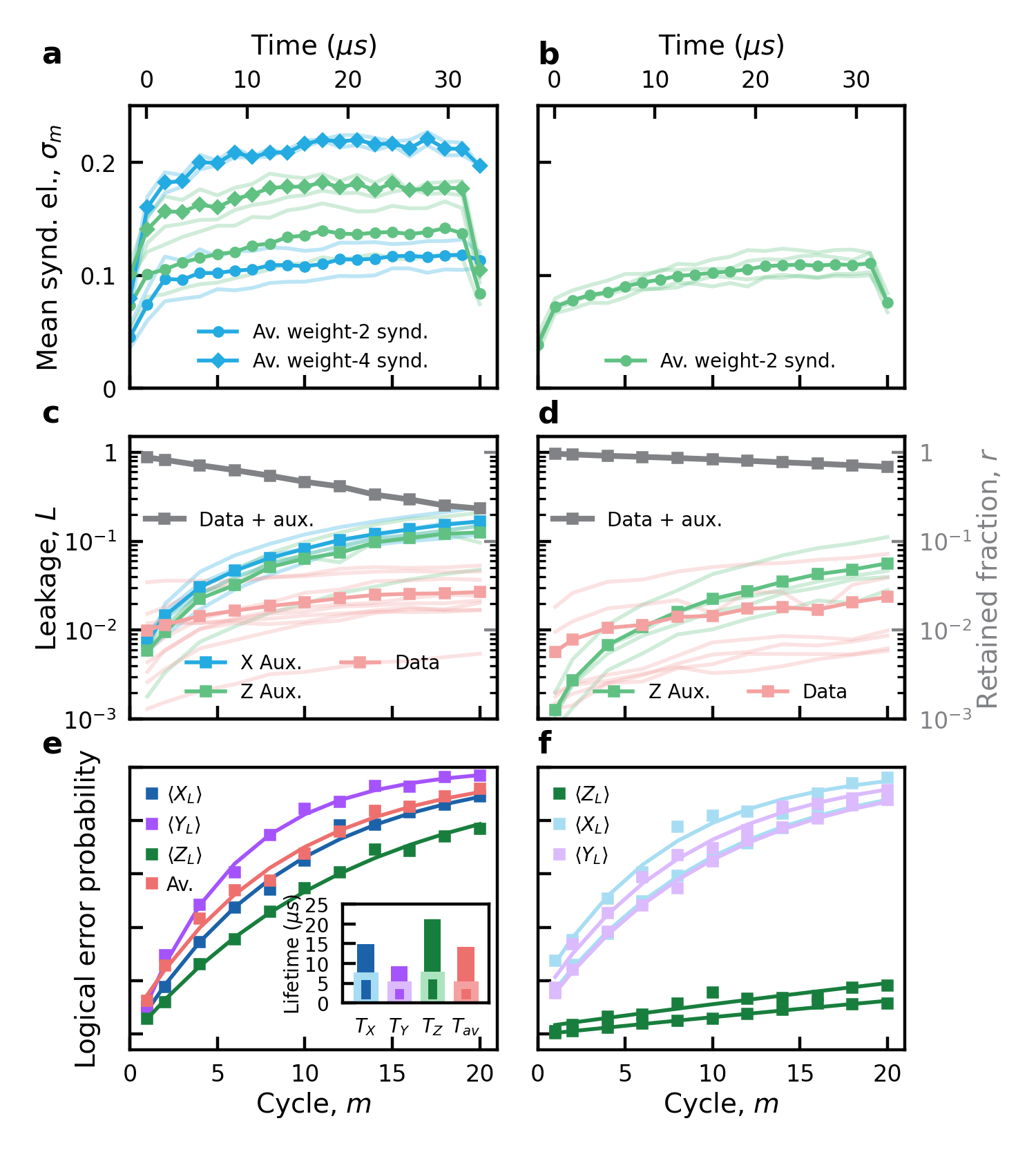}
\caption{\label{fig:SC1}Performance of surface-code and repetition-code experiments. \textbf{a} Mean syndrome elements of the surface-code experiment. The lighter green (blue) lines show $Z$-type ($X$-type) syndrome data, and the darker lines indicate the average over stabilizers with the same weight. \textbf{b} Mean syndrome elements for the repetition codes. The darker line shows the average over all $Z$-syndrome data. \textbf{c} and \textbf{d} Leakage and retention rates for the surface-code and repetition-code experiment. The darker red, blue and green lines indicate the average leakage of data, $X$-type auxiliary, and $Z$-type auxiliary qubits, respectively. The gray line indicates the retained fraction of experimental runs after having postselected on no leakage events. \textbf{e}~(\textbf{f})~Logical error probability for the distance-3 surface code (bit-flip repetition code) qubit observables. 
Inset: extracted raw (wireframe) and decoded lifetimes for different observables of the distance-3 surface-code qubit.
}
\end{figure}

We discard runs where any readout has resulted in a leaked outcome.
For weight-four stabilizers, we find an average syndrome element of 0.182, while for weight-two stabilizers it is 0.114. 
This is expected, as the four-qubit stabilizer syndrome extraction circuit is sensitive to errors in twice as many two-qubit gates.
The initial and final syndromes are flipped by errors in fewer two-qubit gates, which explains the lower average value of these syndrome elements (0.075 and 0.125) compared to bulk syndrome elements (0.153).
The average $Z$-type ($X$-type) syndrome elements are shown in Fig.~\ref{fig:SC1}a as a function of the stabilizer measurement cycle $m$ with green (blue) circles for weight-two syndromes and squares for weight-four syndromes. 

In addition to the differences between weight-two and weight-four syndrome elements, and bulk, initial, and final syndrome elements, bulk syndromes show a gradual increase in their mean values with the cycle number.
One effect contributing to this behavior is biased noise during auxiliary-qubit readout. 
When initializing the logical qubit in the $Z$ ($X$) basis, the $Z$-type ($X$-type) stabilizers are initialized in the $+1$ state. 
Before and during the readout, auxiliary qubits are expected to remain in the ground state, which is not prone to decay errors.
Without auxiliary-qubit reset, errors accumulating over cycles cause the auxiliary qubits to approach an asymptotic population of 0.5 and increase the impact of decay errors. 
Another effect that leads to the observed gradual increase of the syndrome elements is undetected data-qubit leakage accumulation.

Similarly, we extract the mean syndrome elements for the bit-flip codes. 
In this case, we perform fault-tolerant preparation of the $\hat{Z}_\mathrm{L1}\hat{Z}_\mathrm{L2}$ eigenstates: the active data qubits D1, D3, D4, D6, D7, and D9 are initialized in the $\ket{0}^{\otimes 6}$, $\hat{X}_\mathrm{L1}\ket{0}^{\otimes 6}$, $\hat{X}_\mathrm{L2}\ket{0}^{\otimes 6}$, and $\hat{X}_\mathrm{L1}\hat{X}_\mathrm{L2}\ket{0}^{\otimes 6}$ states. 
We perform 20 cycles of syndrome extraction, with the same circuit as used in the second half of the split circuit (Fig.~\ref{fig:gate_sequence}). 
Simultaneously with the auxiliary-qubit readout of the last cycle of syndrome extraction, we read out the data qubits in the $Z$ basis. Similar to the case of the surface code, this yields 19 cycles of bulk syndromes calculated as parities of stabilizer measurement outcomes, and two extra syndromes from the fault-tolerant preparation and readout. We use the same cycle time of $1.66~\mathrm{\mu s}$ as in the distance-three surface code. The average syndrome elements over both bit-flip codes is 0.095, and is shown as a function of the stabilizer measurement cycle $m$ in Fig.~\ref{fig:SC1}b. The higher value of the average weight-two syndrome elements in the distance-three surface code compared to the repetition codes can be attributed to gate errors that occur during the $X$-type syndrome extraction circuit.

Next, we perform a state preservation experiment for the distance-three surface code with varying number of error correction cycles, as in Refs.~\onlinecite{Krinner2022, Acharya2023, Acharya2024a, Ryan-Anderson2021}. 
The $\hat{X}_\mathrm{L}$ ($\hat{Z}_\mathrm{L}$) eigenstates are prepared fault-tolerantly, and after $m$ interleaved cycles of $X$-type and $Z$-type stabilizer measurements, the data qubits are read out in the $X$ ($Z$) basis.  
Additionally, we consider a dataset where the logical qubit is initialized non-fault-tolerantly using the arbitrary state preparation scheme, and logical state tomography is carried out to determine the final state after $m$ interleaved cycles of $X$-type and $Z$-type stabilizer measurements. 
The initialization and readout of arbitrary logical states are described in~App.~\ref{app:arbstate}.

For the bit-flip repetition codes, we perform the logical state preservation experiment with fault-tolerant preparation of the $\hat{Z}_\mathrm{L1}\hat{Z}_\mathrm{L2}$ eigenstates followed by $m$ cycles of $Z$-type syndrome extraction and data-qubit readout in the $Z$ basis. 
Additionally, we perform experiments preparing the data qubits in the $\ket{+}^{\otimes6}$, $\hat{Z}_\mathrm{L1}\ket{+}^{\otimes6}$, $\hat{Z}_\mathrm{L2}\ket{+}^{\otimes6}$, and $\hat{Z}_\mathrm{L1}\hat{Z}_\mathrm{L2}\ket{+}^{\otimes6}$ states, and final data qubit readout in the $X$ basis for determining the error rate in the $X$ observables. 
In this case, the preparation and readout are not fault-tolerant, due to missing $X$-type syndrome extraction. 
For determination of $Y$ observable error rates we prepare separable $\hat{Y}_\mathrm{L1}\hat{Y}_\mathrm{L2}$ eigenstates. 
The active data qubits D1, D3, D4, D6, D7, and D9 are initialized in the $\ket{{+}i{+}{+}i{+}}$, $\hat{Z}_\mathrm{L1}\ket{{+}i{+}{+}i{+}}$, $\hat{Z}_\mathrm{L2}\ket{{+}i{+}{+}i{+}}$, and $\hat{Z}_\mathrm{L1}\hat{Z}_\mathrm{L2}\ket{{+}i{+}{+}i{+}}$ states. Here, $\ket{i}$ stands for the $+1$ eigenstate of the $\hat{Y}$ operator. 
The final data-qubit readout is performed in the $Y$ basis for the middle row of data qubits (D4 and D6), and in the $X$ basis for all other data qubits.

Again, we discard all experimental runs where any one of the qubits has been detected in the leaked state. 
The data qubits are measured only at the end of the gate sequence.
We observe an increase in the average probability of data qubits yielding a leaked state with the number of error correction cycles $m$.
This dependence is shown by the red curves in Fig.~\ref{fig:SC1}c and d. 
For the surface code, the data-qubit leakage probability shows signs of saturation, indicating a finite lifetime of the leaked state, defined as the average number of cycles a qubit remains outside the computational basis.
By fitting exponential models to these dependencies, we obtain per-cycle leakage rates of $2.3(2)\times10^{-3}$ and an average leaked state lifetime of $14\pm2$~cycles for the surface code. 
For the repetition code, the per-cycle leakage rate is $0.1(4)\times10^{-3}$, while the leaked state lifetime exceeds the maximum number of cycles in the experiment and cannot be reliably determined. 

The auxiliary qubits are read out every cycle. 
We investigate the average probability of an auxiliary qubit yielding at least one leaked state readout outcome during $m$ cycles of error correction. 
In the evaluated range of up to $m=20$, these dependencies are linear, showing a per-cycle leakage rate of $8.7(4)\times10^{-3}$ ($7.0(8)\times10^{-3}$) for $X$-type ($Z$-type) auxiliary qubits in the surface code, and $2.9(4)\times10^{-3}$ for the $Z$-type auxiliary qubits in the bit-flip repetition codes.
The green (blue) curves in Fig.~\ref{fig:SC1}c and d show the average probability for $Z$-type ($X$-type) auxiliary qubits yielding at least one leaked state readout outcome in a state preservation experiment with a total of $m$ cycles of error correction. 
The higher auxiliary-qubit leakage compared to data qubits is expected because auxiliary qubits temporarily populate a non-computational state during our CZ gate implementation (see Methods~\ref{subsec:tqg} for details).
Finally, after performing leakage rejection for experimental runs where at least one transmon readout yields the leaked state, we obtain an exponential dependence of retained runs on the number of error correction cycles.
The number of experimental runs remaining after leakage rejection is indicated by the gray curves in Fig.~\ref{fig:SC1}c and d.
Using an exponential model fit, we obtain a per-cycle retention probability of $r_c=0.931$ for the surface code, which is slightly exceeding $r_c=0.921$ cited in Ref.~\onlinecite{Krinner2022}, and $r_c=0.983$ for two simultaneously executed bit-flip repetition codes. 
For the latter, the data-qubit and $Z$-type auxiliary-qubit leakage is reduced due to fewer gate operations contributing to leakage errors.

After leakage rejection, we use syndrome correlation analysis~\cite{Spitz2018} to obtain edge weights for the $Z$ and $X$ decoder graphs. 
The syndrome data are decoded separately for $Z$ and $X$ decoders using minimum-weight perfect matching (MWPM). 
The logical error probability increases with the number of cycles for both the $X$ and $Z$ observables. 
The green (blue) data points in~Fig.~\ref{fig:SC1}e show the average error in the logical $Z_\mathrm{L}$ ($X_\mathrm{L}$) observables calculated from the final data-qubit readout after correcting with the decoder output. 
The blue and green curves show exponential fits to the experimental results. 
From the fits, we obtain per-cycle error probability estimates of $\epsilon_Z=0.078(2)$ for the $Z_\mathrm{L}$ observable and $\epsilon_X=0.111(4)$ for the $X_\mathrm{L}$ observable.

Using the arbitrary state preparation scheme, we demonstrate the preservation of 26 different states on the Bloch sphere (see~App.~\ref{app:arbstate}). 
The average error after $m$ cycles of stabilizer measurements, as determined through logical state tomography, is shown in red in Fig.~\ref{fig:SC1}e (red data points). 
Similar to the $\hat{X}_\mathrm{L}$ and $\hat{Z}_\mathrm{L}$ observables, the error also increases with the number of cycles. 
From the exponential model fit, we obtain an average per-cycle error rate of $\epsilon_\mathrm{av} = 0.117(5)$. 
The $Y_\mathrm{L}$ observable eigenstates exhibit the largest errors, since they are maximally susceptible to both $X$ and $Z$ errors. 
The dependence of the error on the number of cycles for the $Y_\mathrm{L}$ observable is shown in Fig.~\ref{fig:SC1}e as the purple curve. 
The per-cycle error rate for the $Y_\mathrm{L}$ observable is $\epsilon_Y=0.179(6)$. 
The ratios of the cycle time and and the per-cycle error probabilities yield the effective coherence times $T_X, T_Y, T_Z$, and $T_\mathrm{av}$. 
The effective coherence times of the error-corrected logical qubits are significantly higher than those of non-corrected logical qubits for all prepared states.
The comparison between error-corrected and non-error-corrected coherence times is shown in the inset of Fig.~\ref{fig:SC1}e.

For the repetition codes, decoding is only available for $Z$ observables. 
We use a correlation analysis procedure to obtain a decoder graph for the $Z$ syndromes, similar to the surface-code experiment. 
The logical error probability in $\hat{Z}_\mathrm{L1}$ ($\hat{Z}_\mathrm{L2}$) increases with the number of cycles $m$, starting with $0.4\times10^{-2}$ ($1.3\times10^{-2}$) after a single error correction cycle, and reaching $11.5\times10^{-2}$ ($18.1\times10^{-2}$) after 20 cycles. 
The green points in Fig.~\ref{fig:SC1}f show the error probabilities in the $Z$ observable outcomes when decoding with MWPM.
From an exponential fit, we obtain per-cycle error rates of $\epsilon_\mathrm{Z1} = 0.0068(3)$ and $\epsilon_\mathrm{Z2} = 0.0093(11)$ for the two logical qubits.
The $X$ and $Y$ observable outcomes cannot be corrected for phase errors. 
Nevertheless, the raw (uncorrected) outcome values can be evaluated, and the errors in the uncorrected logical observables show similar exponential dependencies, as shown by the light blue and purple points in Fig.~\ref{fig:SC1}f.
Exponential fits to the data yield per-cycle error rates $\epsilon_{X1} = 0.101(2)$, $\epsilon_{X2} = 0.140(8)$, $\epsilon_{Y1} = 0.099(2)$, and $\epsilon_{Y2} = 0.118(5)$. 
These error rates are noticeably worse than those for the $Z$ observables but are only slightly better than the per-cycle error rates for the error-corrected $X$ and $Y$ observables in the distance-three surface code. 
This behavior is expected in the near-threshold regime, where increasing the code distance does neither significantly improve nor degrade the performance of the code.

\section{Distance-One Implementation of Bell-State Preparation Protocol}
\label{app:distance1}

\begin{figure}[t!]
\includegraphics[width=0.5\textwidth]{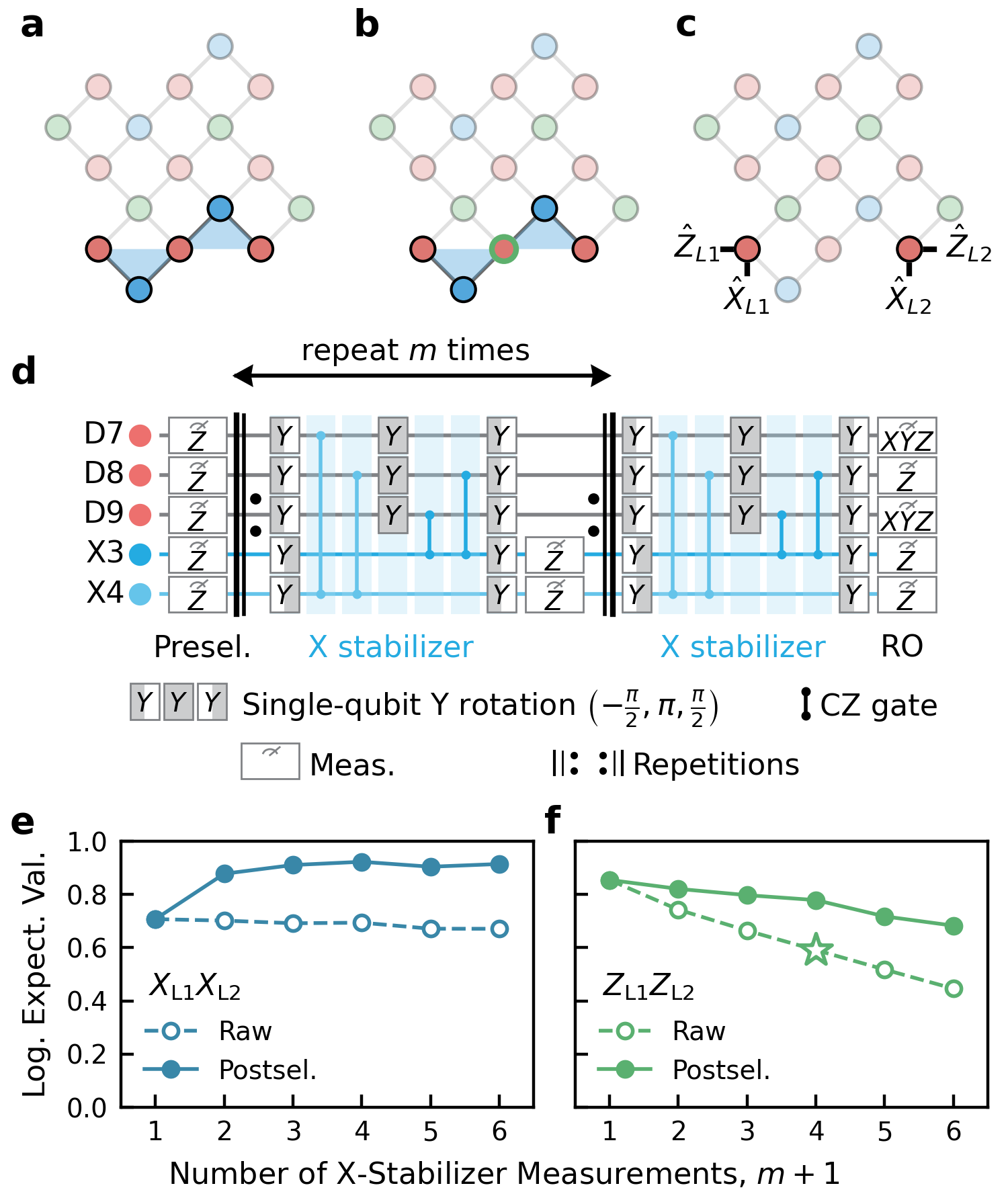}
\centering
\caption{\label{fig:dist-1} Distance-one implementation of the lattice-split protocol. \textbf{a, b, c} Schematic layout of the qubits used in the distance-one protocol. (\textbf{a}) Data qubits D7, D8, D9, used for the Bell-state creation, along with the $X$-stabilizers X3, X4 mediating the entanglement between D7 and D9, are highlighted. (\textbf{b}) Mid-circuit readout of the data qubit D8 indicated as a green circle outline as well as (\textbf{c}) logical operator definitions for the two data qubits sharing entanglement after the operation. \textbf{d} Quantum circuit for the distance-1 protocol. Raw and postselected expectation values of \textbf{e} $X_\mathrm{L1}X_\mathrm{L2}$ and \textbf{f} $Z_\mathrm{L1}Z_\mathrm{L2}$. The star ($\star$) marks the data point corresponding to the same number of repetitions as used in the distance-three implementation in Fig.~\ref{fig:3}a.}
\end{figure}

To compare the bit-flip error-protected protocol with a non-encoded circuit, we implement a distance-one split protocol. 
In this process, we generate a Bell state between two physical qubits within the three-by-three lattice of data qubits. 
This entanglement is created by performing $X$-type stabilizer measurements with a third, central data qubit. 
We select the subset of data and auxiliary qubits exhibiting the best performance. 
In this encoded protocol, all three data qubits, D7, D8, and D9, are initialized in the $\ket{0}$ state, followed by $X$-stabilizer measurements on qubits X3 and X4 to distribute the entanglement, as illustrated in Fig.~\ref{fig:dist-1}a. 
After $1\leq m+1\leq6$ cycles of syndrome extraction, we read out the central data qubit D8 in the $Z$ basis (see Fig.~\ref{fig:dist-1}b). 
For the remaining two physical qubits, we perform a tomographic readout (see Fig.~\ref{fig:dist-1}c), updating the Pauli frame based on the readout of qubit D8 and the two $X$-stabilizer measurement outcomes. 
The quantum circuit for this protocol is shown in Fig.~\ref{fig:dist-1}d.

When plotting the leakage-rejected and Pauli-frame-updated expectation value of $X_\mathrm{L1}X_\mathrm{L2}$ as a function of the number of $X$-syndrome extraction cycles $m$, we observe a value close to 0.7, independent of the number of cycles, as shown by the empty circles in Fig.~\ref{fig:dist-1}e. 
This is because, for the Pauli-frame update of $X_\mathrm{L1}X_\mathrm{L2}$ only the most recent $X$-stabilizer values are used, and not those from earlier cycles. 
By postselecting runs with consistent $X$-stabilizer values across all cycles, we operate this distance-one implementation in a phase-flip error detection mode. 
For the postselected data, we find that, when only one round of stabilizer extraction is performed, postselection does not enhance the observable value. 
This is because the $X$-stabilizer values for qubits initialized in $\ket{0}$ are not predefined and become fixed only after the first round of stabilizer measurements. 
As a result, the initial round cannot be used for postselection. 
However, in subsequent cycles, we observe improvements in the $X_\mathrm{L1}X_\mathrm{L2}$ observable value by postselecting on consistent $X$-stabilizer measurements.

Similarly, we track the $Z_\mathrm{L1}Z_\mathrm{L2}$ observable as a function of the number of $X$-stabilizer measurement cycles, $m$. 
In this case, we observe a reduction in the raw observable value with $m$, primarily due to energy relaxation of the data qubits (empty circles in Fig.~\ref{fig:dist-1}f). 
Since this distance-one implementation lacks $Z$-stabilizer measurements, we are neither able to correct nor detect bit-flip errors. 
For $m+\nolinebreak1=4$, corresponding to the same number of cycles used in the distance-three implementation, we find an observable value of $0.591(8)$ for $Z_\mathrm{L1}Z_\mathrm{L2}$, indicated by an open star in Fig.~\ref{fig:dist-1}f. 
We also observe a modest improvement on the $Z_\mathrm{L1}Z_\mathrm{L2}$ observable when postselecting on consistent $X$-stabilizer measurements, which we attribute to $Y$ errors. $Y$ errors can be detected by $X$ syndromes, but affect both $X$ and $Z$ observables.

\section{Error Correction for Tomographic Readouts\label{app:decoding}}

For decoding, we assume a Pauli circuit noise model. 
In this model, during the Bell state preparation and the measurement of the $Z_{\mathrm{L}1}Z_{\mathrm{L}2}$ observable, all errors that flip the same combination of syndromes have the same effect on the logical observable. 
This property underpins the fault tolerance of the circuit.

Any error affects at most two $X$ syndrome elements and at most two $Z$ syndrome elements. 
Any $Y$ error, i.e., an error that flips both $X$ and $Z$ syndrome elements, can be represented as two separate errors: one affecting only $X$ syndromes and the other affecting only $Z$ syndromes. 
These decomposed errors have the same impact on the observables as the original error. 
Leveraging this separability property, we decompose the decoding problem into two smaller subproblems: decoding bit-flip errors, also referred to as $X$ errors, using $Z$ syndrome data and decoding phase-flip ($Z$) errors using $X$ syndrome data. 
This approach simplifies the decoding process, but it does so at the cost of disregarding correlations between $X$ and $Z$ errors that are due to the decomposition.

The decoding problem can be formulated as finding the most likely combination of errors that matches the observed syndrome elements. 
This is achieved by using graph representations for the $X$ and $Z$ syndrome elements. 
In the syndrome graph, each syndrome element corresponds to a node, while edges represent error mechanisms that flip the two connected syndrome elements. 
To handle errors that flip only a single syndrome element, a special “boundary” node is introduced, which is not associated with any measured syndrome. 
Edges connected to the boundary node are referred to as boundary edges.
Each edge is assigned a weight $w$, which is related to the probability $p$ of the error mechanism causing a flip of the associated syndrome elements. 
The relation between weight and probability is $w=-\log(p/(1-p))$.
Within this graph representation, the decoding problem reduces to solving the minimum-weight perfect matching (MWPM) problem, which we address using the blossom algorithm~\cite{Edwards1975}, implemented in the pymatching package~\cite{Higgott2022}.

For quantum state tomography, all edges in the $Z$ ($X$) decoding graph are classified based on whether the associated errors flip the $Z_{\mathrm{L}1}$ or $Z_{\mathrm{L}2}$ ($X_{\mathrm{L}1}$ or $X_{\mathrm{L}2}$) logical observables. 
Any two-logical-qubit observable can be expressed as a product of these four operators. 
Using this representation, along with the results of the MWPM algorithm, we determine whether a correction is required for the raw measurement outcome.

We find that in the circuit noise model, all errors associated with the same edge in the $Z$ matching graph have the same effect on the logical observables for fault-tolerant preparation of $\hat{Z}_\mathrm{L}$ eigenstates. 
For the most likely errors, the mapping between edges and their effects on the logical observables is shown in Fig.~\ref{fig:3}b.
However, if the final readout of one of the logical qubits is performed outside the $Z$ basis, the final set of syndrome elements, $\sigma_5^{\mathrm{Z}i}$, becomes non-deterministic, i.e., cannot be computed because the final stabilizer value is unavailable.
In this case, the syndromes from the final cycle are not used.

\begin{figure}[t!]
\includegraphics[width=0.22\textwidth]{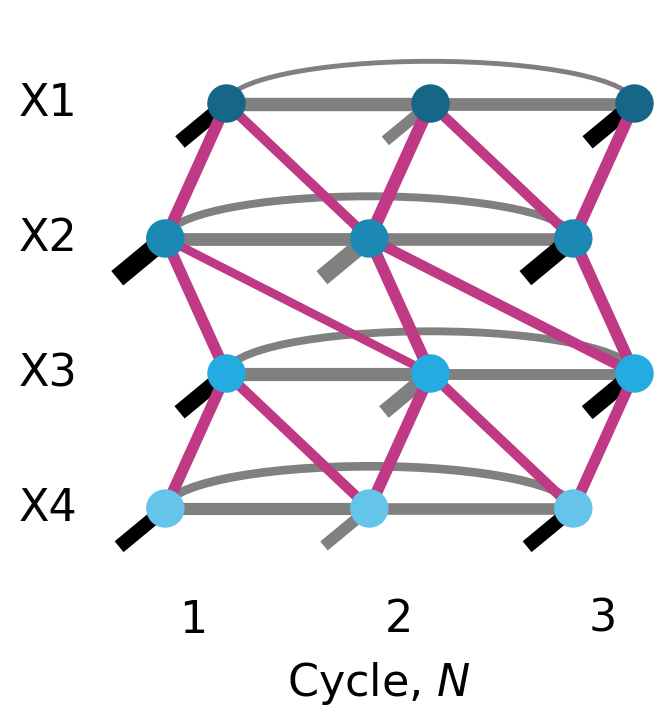}
\caption{\label{fig:F1}Experimentally extracted decoder matching graph weights for $X$ syndromes and their impact on logical observables. 
Gray edges correspond to errors that flip neither $X_\mathrm{L1}$ nor $X_\mathrm{L2}$. 
Purple edges indicate errors that flip both $X_\mathrm{L1}$ and $X_\mathrm{L2}$. 
Ambiguous edges associated with multiple errors with different effects on the logical observables are shown in black.
}
\end{figure}

For the $X$-type syndrome data, the circuit noise model yields three classes of edges: those corresponding to errors that do not flip any logical observables, those corresponding to errors that flip both $\hat{X}_{\mathrm{L}1}$ and $\hat{X}_{\mathrm{L}2}$, and ambiguous boundary edges.
These edge classes are shown with different colors in Fig.~\ref{fig:F1} together with the experimentally extracted edge weights \cite{Spitz2018}. 
Ambiguous boundary edges arise due to the absence of $X$-type stabilizer data before the first round and after the split operation.
Since there are no edges that are unambiguously associated with a flip of either $\hat{X}_{\mathrm{L}1}$ or $\hat{X}_{\mathrm{L}2}$, but not both, there are no phase-flip errors that can be corrected for the $X_{\mathrm{L}1}X_{\mathrm{L}2}$ and $Y_{\mathrm{L}1}Y_{\mathrm{L}2}$ observables. 
Therefore, for the Bell state, $X$ syndrome decoding does not provide any improvement to the state fidelity. 

For input states with non-zero expectation values of the $X_\mathrm{L}$ or $Y_\mathrm{L}$ observables, the split operation yields non-trivial expectation values for the $X_{\mathrm{L}1}I_{\mathrm{L}2}$, $I_{\mathrm{L}1}X_{\mathrm{L}2}$, $Y_{\mathrm{L}1}Z_{\mathrm{L}2}$, and $Z_{\mathrm{L}1}Y_{\mathrm{L}2}$ observables, which are unambiguously flipped by errors corresponding to purple edges shown in Fig.~\ref{fig:F1}. 
These edges are associated with phase-flip errors occurring during the operation of the distance-three surface code and can be corrected.

For the arbitrary state preparation scheme, described in detail in App.~\ref{app:arbstate}, the first round of weight-four syndromes is non-deterministic, and as a result unavailable, while all the weight-two syndromes are deterministic (including $\sigma_0^{\mathrm{X}1}$ and $\sigma_0^{\mathrm{X}4}$).
This results in a loss of fault-tolerance due to ambiguous boundary edges and undetectable errors during initialization. 
The appearance of these ambiguous boundary edges and undetectable errors arises from the loss of the deterministic weight-four stabilizers ($\sigma_0^{\mathrm{Z}2}$ and $\sigma_0^{\mathrm{Z}3}$) in the first round. 
In the current work, we do not attempt to modify the syndrome graph, but instead postselect on runs where no non-trivial syndromes are detected in the initial cycle.

Summarizing, in this work, we separately correct phase-flip and bit-flip errors using the $X$ and $Z$ syndrome graphs, respectively, wherever possible. 
For the arbitrary state preparation, we do not correct errors but discard datasets where non-trivial syndromes are triggered during the preparation cycle.

\section{Simulation}
\label{app:Simulation}

\begin{figure}[ht!]
\includegraphics[width=0.5\textwidth]{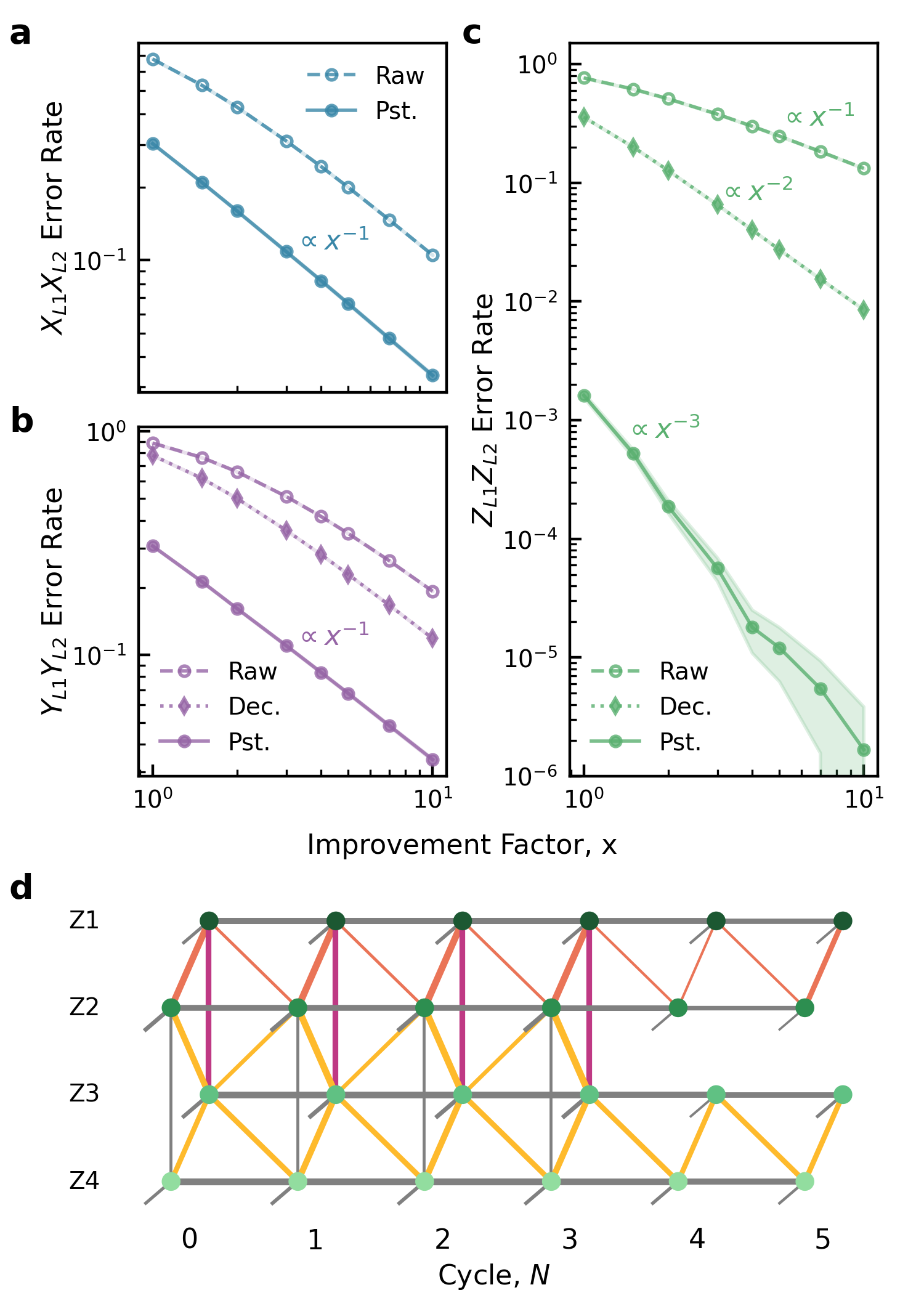}
\caption{\label{fig:sim}Results of the stabilizer simulation of the lattice-split protocol. \textbf{a, b, c} Raw, decoded and postselected expectation values of (\textbf{a})~$X_\mathrm{L1}X_\mathrm{L2}$, (\textbf{b})~$Y_\mathrm{L1}Y_\mathrm{L2}$, and (\textbf{c})~$Z_\mathrm{L1}Z_\mathrm{L2}$ as a function of error rate improvement factor that scales down uniformly all modeled physical noise sources. The shaded regions indicate statistical uncertainty from Monte Carlo sampling. \textbf{d}~Matching graph derived from the error model, with circles corresponding to syndrome elements and edges to possible errors, with their probability encoded in the thickness of the edge. Red (yellow) edges cause a logical error on the $Z_\mathrm{L1}$ ($Z_\mathrm{L2}$) observable. Gray edges  cause no logical observable error. Purple edges cause errors on both $Z_\mathrm{L1}$ and $Z_\mathrm{L2}$.}
\end{figure}

We use Monte Carlo sampling to estimate logical error rates for the gate sequence used in the experiment, shown in Fig.~\ref{fig:gate_sequence}, using the stim~\cite{Gidney2021} and pymatching~\cite{Higgott2023} Python packages. 
Our noise model incorporates both single- and two-qubit depolarizing channels for the single- and two-qubit gates. 
The depolarization probabilities are determined from error rates measured via randomized benchmarking for single-qubit gates (see Methods Section~\ref{subsec:sqg}) and interleaved randomized benchmarking for two-qubit gates (see Methods Section~\ref{subsec:tqg}), performed individually for each qubit and qubit pair.

\begin{figure*}[t!]
\includegraphics[width=\textwidth]{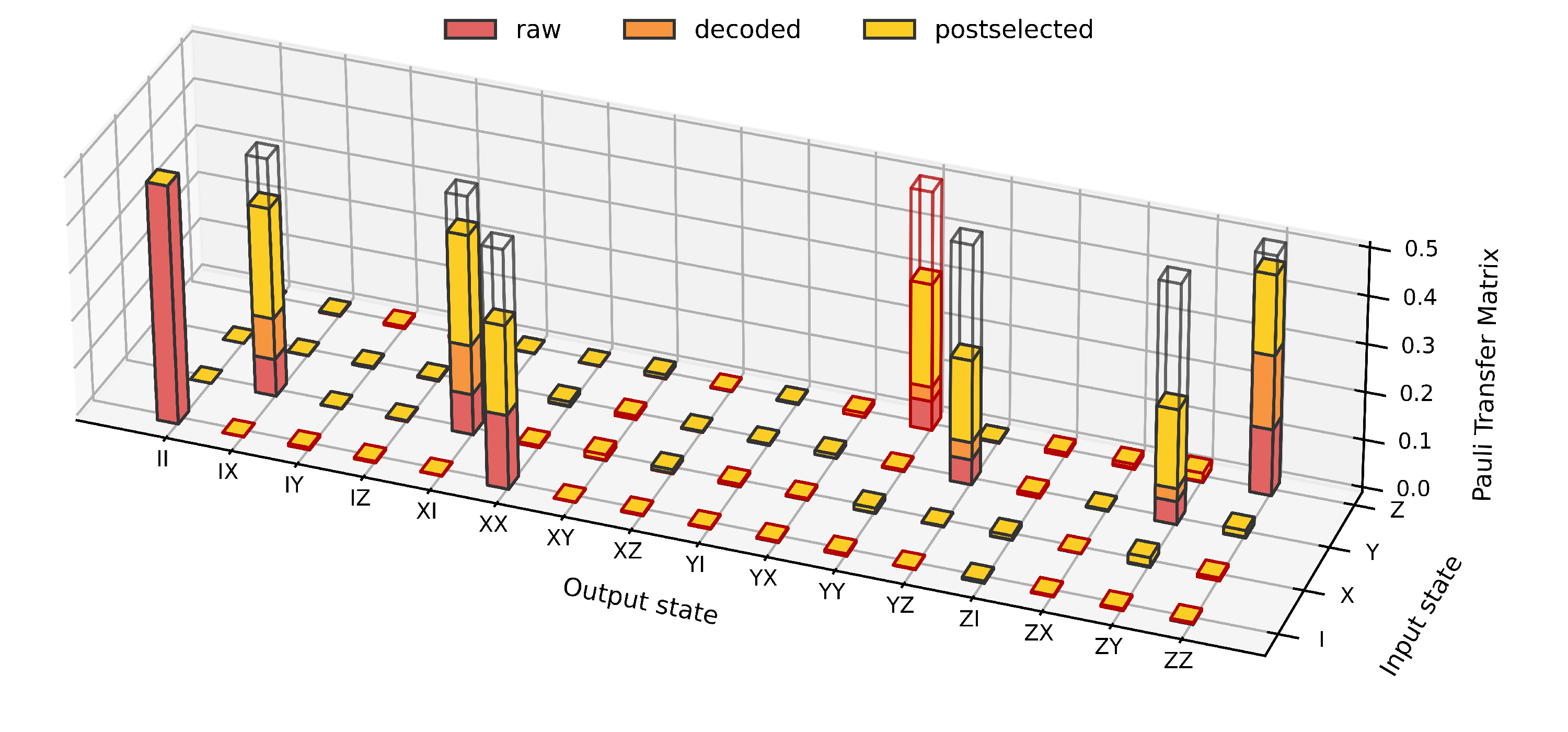}
\centering
\caption{\label{fig:E5}Simulated Pauli transfer matrix (PTM) of the split operation reconstructed from simulated raw, decoded and postselected logical observable outcomes. 
The designations match those of Fig.~\ref{fig:5}.}
\end{figure*}

Readout errors are simulated by applying a Pauli-$X$ gate before a perfect measurement, with the probability determined by the experimentally characterized two-state readout assignment error. 
Initialization errors are modeled as an imperfect readout operation applied to an ideal ground state. 
At each step of the circuit, if a qubit is not involved in an active operation, a Pauli error channel based on the coherence properties of the qubit is applied. 
The probabilities for this error channel are given by $p_i = 1-e^{-t/T_i}$, with $T_X=T_Y=4T_1$, and $T_Z=T_\varphi$. 
The energy relaxation time $T_1$ and Hahn echo time $T_{2,E}$ are experimentally determined for each qubit, and the pure dephasing rate is computed as
\begin{align*}
    \frac{1}{T_\varphi} = \frac{1}{T_{2,E}} - \frac{1}{2T_1}.
\end{align*}
The value of $t$ corresponds to the gate duration as used in the experimental implementation. 
We use Ramsey echo time, rather than the decoherence time $T_2^*$, as our experimental implementation incorporates dynamical decoupling pulses. 
Other known noise sources, such as drive crosstalk, flux crosstalk, measurement-induced dephasing, and leakage, are not included in the noise modeling and simulations.

We sample the $X_\mathrm{L1}X_\mathrm{L2}$, $Y_\mathrm{L1}Y_\mathrm{L2}$, and $Z_\mathrm{L1}Z_\mathrm{L2}$ logical observables by multiplying the readout outcome of appropriately chosen data qubits. 
To investigate, in a coarse way, the effect of future anticipated experimental improvements of operations, we apply an improvement factor of $x$, scaling all Pauli error probabilities by a constant factor of $\frac{1}{x}$. 
For each improvement factor and logical observable, we sample $10^7$ experiment outcomes, from which we calculate the corresponding logical operator expectation values. 
For all three observables, the raw logical observable error rates scale linearly with the improvement factor (see Fig.~\ref{fig:sim}a, b, and c). 
By discarding simulated outcomes where one or more syndrome elements are non-zero, we obtain postselected observable values. 
For the $Z_\mathrm{L1}Z_\mathrm{L2}$ observable, we find a cubic scaling of the postselected error rate with the improvement factor compared to the linear scaling of the raw error rate.
This is consistent with the expected behavior of a protocol where the smallest undetectable fault weight causing a logical error is three. 
For the other observables, while postselection improves the error rates, the scaling remains linear.

The matching graph weights for simulation-based decoding are derived from the probabilities within the detector error model of the stim circuit with the specified noise model. 
The resulting graph is qualitatively similar to the experimental one shown in Figure \ref{fig:3}b. 
Comparing the connectivity of both graphs, we recover the absence of significant correlations between the syndromes of the two repetition codes after the split operation, as can be seen by the missing edges connecting the stabilizer qubits Z1, Z2 with Z3, Z4 for cycles $m=4,5$, see Fig.~\ref{fig:sim}d. 
A key difference is the absence of the weight-two time-like correlations, which are, on the other hand, present in the experimental decoding graph.
In our gate sequence, which does not feature auxiliary-qubit reset after readout, such correlations can be caused by readout misclassification errors or data-qubit leakage. 
Both of these error mechanisms are absent in our simulation. 
As with the experimental data, we apply minimum-weight perfect matching decoding using the matching graph. 
This yields a quadratic error scaling for the decoded $Z_\mathrm{L1}Z_\mathrm{L2}$ observable, as expected for a fault-tolerant operation. 
We also observe some improvement in the decoded error rate of the $Y_\mathrm{L1}Y_\mathrm{L2}$ observable, compared to the raw error rate, due to the correction of bit-flip errors during the protocol. 

The simulated operator values for $x=1$ are represented by the wireframes in Fig.~\ref{fig:3}b. 
We quantify the agreement between experimental and simulated data by computing the norm-based fidelity~\cite{Liang2019} between the simulated and experimentally extracted logical density matrices as
\begin{align*}
    \mathcal{F}_\text{overlap} = \operatorname{Tr}\left[\hat{\rho}_\mathrm{sim}\hat{\rho}_\mathrm{exp}\right]/\operatorname{max}\left[\operatorname{Tr}\hat{\rho}_\mathrm{sim}^2, \operatorname{Tr}\hat{\rho}_\mathrm{exp}^2\right]
\end{align*}
and find an average value of $\mathcal{F}_\text{overlap} = 96\%$ for the raw, decoded and postselected data. 
While the simulation yields good quantitative agreement overall, there are some systematic differences among the operators. 
Specifically, experimental data for $X_\mathrm{L1}X_\mathrm{L2}$ and $Y_\mathrm{L1}Y_\mathrm{L2}$ show, on average, lower expectation values compared to the ones expected from simulations, while $Z_\mathrm{L1}Z_\mathrm{L2}$ aligns more closely with experimental values. 
We attribute this discrepancy to coherent phase errors, which impact only the $X_\mathrm{L1}X_\mathrm{L2}$ and $Y_\mathrm{L1}Y_\mathrm{L2}$ observables but are not accounted for in our noise model.

We also simulate the Pauli transfer matrix reconstruction experiment using the same gate error model as above but with an initial state preparation circuit adapted to match the arbitrary state preparation protocol. 
Following the experimental post-processing procedure, runs with errors detected during the first cycle of stabilizer measurements are discarded. 
The Pauli transfer matrix reconstructed from the simulated data is shown in Fig.~\ref{fig:E5}. 
We observe qualitative agreement with the experimentally extracted PTM, except for the absence of entries corresponding to coherent over-rotations (see App.~\ref{app:PhaseRotation}), which are not included in our error model.

\section{Coherent Rotation Quantification and Correction\label{app:PhaseRotation}}

\begin{figure}[t!]
\includegraphics[width=0.5\textwidth]{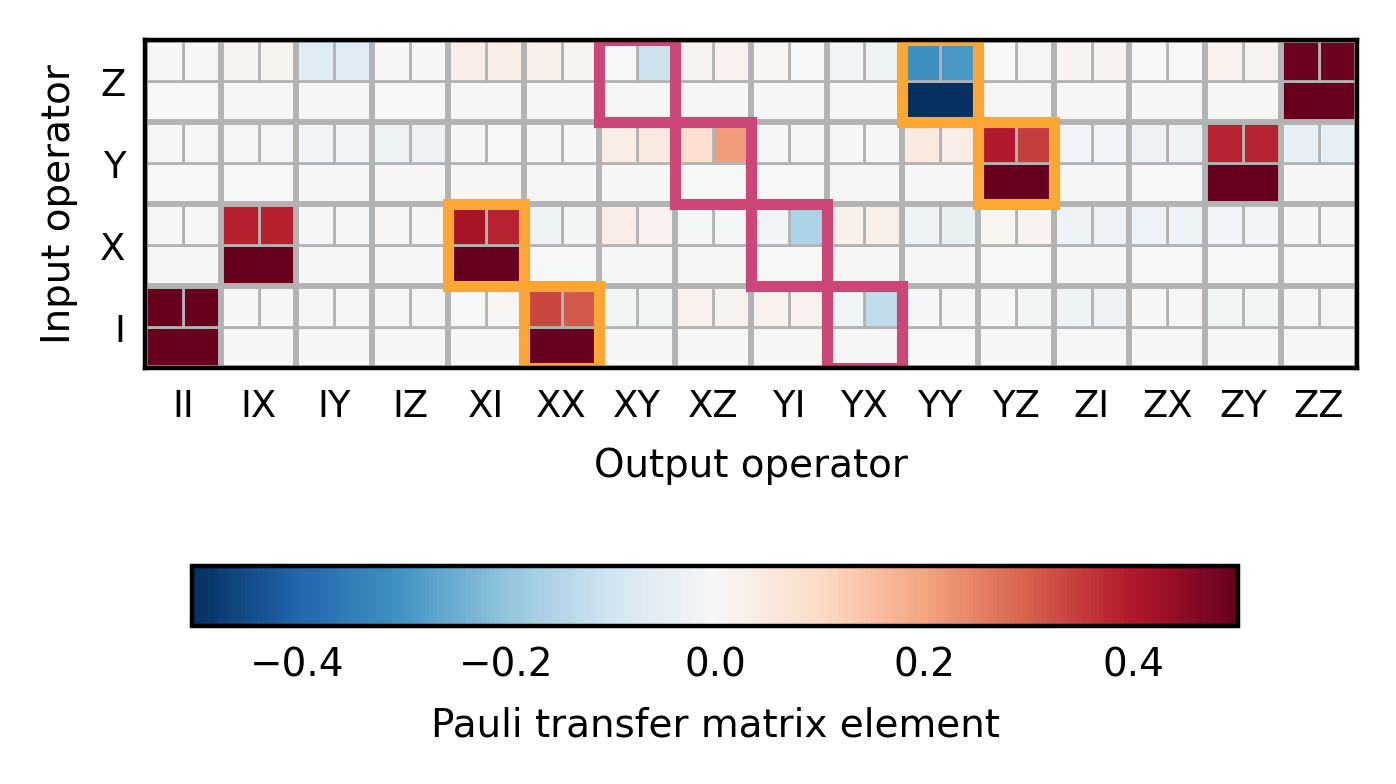}
\caption{\label{fig:SF1}Postselected Pauli transfer matrix (PTM) for the split operation with and without correction of the phase error on the first logical qubit. Each element in the PTM is color-coded. The bottom part of each cell shows the PTM entry corresponding to an ideal split operation, the left upper part of the cell corresponds to the corrected entry, and the right upper part of the cell corresponds to the uncorrected entry. The entries most affected by the rotation are indicated by colored wireframes, orange for entries whose magnitude is reduced due to the rotation, and purple for entries whose magnitude is increased due to the rotation.
}
\end{figure}

In this section, we discuss the single-logical-qubit phase rotation observed in the logical state tomography (Fig.~\ref{fig:3}c and d) and logical process tomography (Fig.~\ref{fig:5}).
Applying a $0.11\pi$ rotation around the $Z$ axis on the first logical qubit improves the fidelity of the Bell state from~0.780(6) to~0.788(6), and the fidelity of the process from~0.78(3) to~0.81(3), after postselection for no detected errors. 
In Fig.~\ref{fig:SF1} we present the Pauli transfer matrix with and without the phase rotation correction. 
The unintended phase rotation has the strongest effect on the highlighted matrix elements.
For an ideal split operation without phase errors, the entries highlighted in purple should be zero.
The effect of the rotation is that these entries increase in magnitude, while entries with the same input operator outlined in orange decrease.
After applying the correction, the absolute values of the unintended entries outlined in purple are reduced. 
The deviations of the remaining entries in the PTM from the expected value of zero are within a 95\% confidence interval. After postselection for no detected errors, only 142 out of $3\,062$ runs per prepared state and measurement basis remain.

For the logical Bell state, the final state of the data qubits is, up to a Pauli-frame update, a GHZ state: $\frac{1}{\sqrt{2}}\left(\ket{0}^{\otimes 6}+\ket{1}^{\otimes 6}\right)$. 
The phase of the non-zero off-diagonal matrix element in Fig.~\ref{fig:3}c is the sum of the individual single-qubit phase rotations of the data qubits. 
To identify the source of the phase error, we perform correlation analysis between matched edges in the $Z$ syndrome graph and the measurement outcomes of the $X_\mathrm{L1}Y_\mathrm{L2}$ and $Y_\mathrm{L1}X_\mathrm{L2}$ observable. 
These observables are affected by bit-flip errors, but only those that flip one of the constituent logical qubits.
The analysis shows that the bit-flip errors affect the logical observables as shown in Fig.~\ref{fig:3}b, i.e., $X_\mathrm{L1}Y_\mathrm{L2}$ is flipped by bit-flip errors that only affect the second logical qubit, and $Y_\mathrm{L1}X_\mathrm{L2}$ by those that only affect the first logical qubit. 
This behavior can be expected, e.g., if the frequency of a data qubit is incorrectly calibrated or fluctuating as a result of coupling to a two-level system defect. 
While dynamical decoupling pulses prevent the accumulation of such errors over time, they can still lead to increased detected error rates and coherent rotations for observables whose preparation and readout are not fault-tolerant.

\section{Arbitrary State Preparation and Tomography of a Logical Qubit}
\label{app:arbstate}

\begin{figure}[t!]
\includegraphics[width=0.5\textwidth]{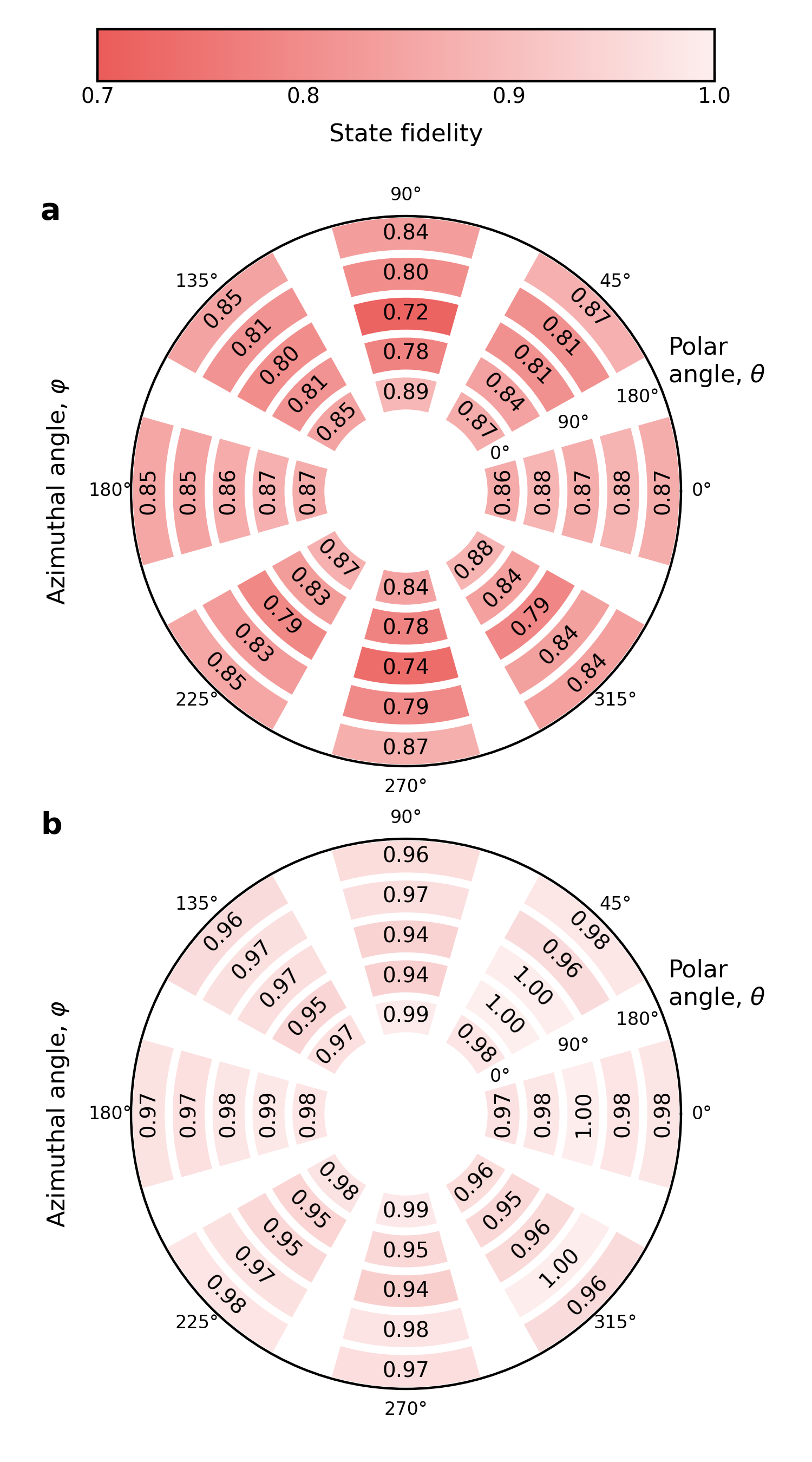}
\centering
\caption{\label{fig:arb_state_fid} Arbitrary state preparation and readout performance figures \textbf{a} with postselection only on leakage events and \textbf{b} with postselection both on leakage and non-trivial syndromes.}
\end{figure}

The surface code enables fault-tolerant preparation of the logical $\hat{X}_\mathrm{L}$ and $\hat{Z}_\mathrm{L}$ eigenstates, as described in App.~\ref{app:performance}. 
This is possible because errors affecting these states can be corrected using either only $X$-type or only $Z$-type syndrome data, respectively.  
For arbitrary logical states, fault-tolerant preparation requires distillation methods~\cite{Bravyi2005}.
Nevertheless, it is possible to prepare an arbitrary state of the distance-three surface code non-fault-tolerantly. Here, we follow the approach introduced in~\cite{Ye2023a, Li2015l, Lao2022}, which yields an incomplete subset of the stabilizers after the preparation. 
The arbitrary state is injected into the central data qubit (D5).
The other data qubits are initialized as illustrated in Fig.~\ref{fig:4}a, ensuring that all weight-two stabilizers are predefined. 
This approach allows us to compute four out of eight syndrome elements for the initial cycle of error correction.
These syndrome elements do not account for all possible single-qubit and two-qubit gate errors, such as those affecting D5. 
Moreover, these syndromes are insufficient to determine the effect the errors will have on the logical observables, rendering correction impossible.
Nevertheless, by postselecting runs where the first cycle of stabilizer measurements yields the expected outcomes for the weight-two stabilizers, we are able to detect some of the possible errors during state preparation.
From a code deformation perspective this preparation can be understood as adding eight data qubits and stabilizers to a surface-code patch consisting only of D5, which encodes the original logical qubit.

Similarly, by applying code deformation to the distance-three surface-code qubit in reverse, we can perform a logical-qubit measurement in an arbitrary basis. 
All data qubits, except for D5, are read out in the basis of the weight-two stabilizer they are included in. 
D5 can be read out in any basis. In our experiment, we use logical quantum state tomography, and the arbitrary-basis measurement is only needed for the $\hat{Y}_\mathrm{L}$ operator, since $\hat{X}_\mathrm{L}$ and $\hat{Z}_\mathrm{L}$ are measured in a fault-tolerant manner. 
The $\hat{Y}_\mathrm{L}$ operator is defined as the product of $\hat{Z}_\mathrm{L}$ and $\hat{X}_\mathrm{L}$, which in terms of data qubits is $\hat{X}_2\hat{Z}_4\hat{Y}_5\hat{Z}_6\hat{X}_8$. 
To improve the readout fidelity, we discard runs where one of the stabilizers Z1, Z4, X1, or X4 extracted from the final data-qubit readout does not match the final round of stabilizer measurements.

By combining these two methods, we can assess our ability to herald arbitrary logical states and characterize their fidelity using logical quantum state tomography. 
To achieve this, we prepare the central data qubit (D5) in a desired arbitrary state $\ket{\psi}$, parameterized by the polar and azimuthal angles introduced in Fig.~\ref{fig:4}b, with the remaining data qubits initialized as described above. 
We then perform a complete round of $Z$- and $X$-type stabilizer measurements to initialize the distance-three surface-code qubit in the arbitrary state. 
Finally, by measuring the data qubits in the three logical bases $\hat{X}_\mathrm{L}$, $\hat{Y}_\mathrm{L}$, and $\hat{Z}_\mathrm{L}$, as described earlier, we compute the fidelity according to
\begin{align}
    \mathcal{F}_\psi = \frac{1}{2}\left(1+\langle \hat{X}_\mathrm{L}\rangle\langle\hat{\sigma}_x\rangle_\psi+\langle \hat{Y}_\mathrm{L}\rangle\langle\hat{\sigma}_y\rangle_\psi+\langle \hat{Z}_\mathrm{L}\rangle\langle\hat{\sigma}_z\rangle_\psi\right),
\end{align}
where $0 \leq \langle\hat{X}_\mathrm{L}\rangle, \langle\hat{Y}_\mathrm{L}\rangle, \langle\hat{Z}_\mathrm{L}\rangle \leq 1$ represent the logical readout results in the respective basis. 
Fidelities exceeding unity are clipped to ensure positive semi-definiteness of the density matrix.
The fidelities obtained using this method are shown in Fig.~\ref{fig:arb_state_fid}, both for the case of leakage rejection only (a) and the case of leakage rejection combined with postselection on no detected syndrome events (b). 
Without syndrome postselection\-, we achieve an average fidelity of $83.5\%$, which improves to $97.0\%$ when error detection is applied. 
We observe that the lowest fidelities occur for the $\hat{Y}_\mathrm{L}$ eigenstates, specifically for $\theta = 90^{\circ}$ and $\varphi = \pm 90^{\circ}$, where both phase-flip and bit-flip errors contribute to the infidelity. 
Averaging over all angles, out of the initial 3~322 repetitions, 3~044 remain after leakage rejection, and about 1~577 remain after postselection on no syndrome events.

\bibliographystyle{apsrev4-2-title-etal}
\bibliography{QudevRefDB}

\end{document}